\documentclass[12pt]{article}
\usepackage{epsfig,amsfonts,amssymb}
\usepackage{hyperref}

\usepackage{cite}
\usepackage{cite}

\usepackage{comment}

\usepackage{amssymb, latexsym, amsmath, amsthm}
\usepackage[left=2cm, right=2cm, top=2cm]{geometry}
\usepackage{float}
\usepackage{caption}
\usepackage{subcaption}
\usepackage{enumitem}
\usepackage{bbm}
\usepackage{longtable}
\usepackage{fullpage} 
\usepackage{parskip} 
\usepackage{hyperref}
\usepackage{enumitem}
\usepackage{mathtools}

\theoremstyle{plain}
\newtheorem{thm}{Theorem}[section]

\newtheorem{lemma}[thm]{Lemma}

\theoremstyle{definition}

\theoremstyle{remark}

\usepackage{url}
\usepackage{geometry}
\usepackage{ragged2e}

\def\ZZZ{{\mathbb Z}}
\def\RRR{{\hbox{ R\kern-2.4mm R}}}
\def\CCC{{\hbox{ C\kern-2.0mm C}}}
\def\zzz{{\hbox{z\kern-1mm z}}}

\newcommand{\qeq}{{\hbox{=\kern-2.3mm ? \kern.5mm }}}
\renewcommand{\qeq}{=}

\newcommand{\AAA}{{\cal A}}
\newcommand{\GG}{{\cal G}}

\newcommand{\MM}{{\cal M}}

\newcommand{\wt}{\widetilde}

\newcommand{\be}{\begin{equation}}
\newcommand{\ee}{\end{equation}}
\newcommand{\ben}{\begin{eqnarray}\displaystyle}
\newcommand{\een}{\end{eqnarray}}

\newcommand{\refb}[1]{(\ref{#1})}
\newcommand{\p}{\partial}
\newcommand{\sectiono}[1]{\section{#1}\setcounter{equation}{0}}

\def\one{{\hbox{ 1\kern-.8mm l}}}
\def\zero{{\hbox{ 0\kern-1.5mm 0}}}

\newcommand{\bea}[1]{\begin{eqnarray}\label{#1} }
\newcommand{\eea}{\end{eqnarray}}

\usepackage{bm}
\usepackage[table]{xcolor}

\def\figone{

\def\JPicScale{0.8}
\ifx\JPicScale\undefined\def\JPicScale{1}\fi
\unitlength \JPicScale mm
\begin{picture}(105,75)(0,0)
\linethickness{0.3mm}
\put(30,50){\line(1,0){50}}
\linethickness{0.3mm}
\put(30,0){\line(0,1){50}}
\linethickness{0.3mm}
\put(30,0){\line(1,0){50}}
\linethickness{0.3mm}
\put(80,0){\line(0,1){50}}
\linethickness{0.3mm}
\multiput(30,50)(0.12,0.12){167}{\line(1,0){0.12}}
\linethickness{0.3mm}
\put(50,70){\line(1,0){50}}
\linethickness{0.3mm}
\multiput(80,50)(0.12,0.12){167}{\line(1,0){0.12}}
\linethickness{0.3mm}
\multiput(80,0)(0.12,0.12){167}{\line(1,0){0.12}}
\linethickness{0.3mm}
\put(100,20){\line(0,1){50}}
\linethickness{0.1mm}
\put(50,20){\line(0,1){50}}
\linethickness{0.1mm}
\multiput(30,0)(0.12,0.12){167}{\line(1,0){0.12}}
\linethickness{0.1mm}
\put(50,20){\line(1,0){50}}
\put(25,50){\makebox(0,0)[cc]{$Q_1$}}

\put(25,0){\makebox(0,0)[cc]{$P_1$}}

\put(85,50){\makebox(0,0)[cc]{$Q_2$}}

\put(85,0){\makebox(0,0)[cc]{$P_2$}}

\put(55,73){\makebox(0,0)[cc]{$Q_4$}}

\put(55,23){\makebox(0,0)[cc]{$P_4$}}

\put(100,73){\makebox(0,0)[cc]{$-Q_3$}}

\put(94,23){\makebox(0,0)[cc]{$-P_3$}}

\end{picture}

}

\def\figone{

\def\JPicScale{0.8}
\ifx\JPicScale\undefined\def\JPicScale{1}\fi
\unitlength \JPicScale mm
\begin{picture}(105,75)(0,0)
\linethickness{0.3mm}
\put(30,50){\line(1,0){50}}
\linethickness{0.3mm}
\put(30,0){\line(0,1){50}}
\linethickness{0.3mm}
\put(30,0){\line(1,0){50}}
\linethickness{0.3mm}
\put(80,0){\line(0,1){50}}
\linethickness{0.3mm}
\multiput(30,50)(0.12,0.12){167}{\line(1,0){0.12}}
\linethickness{0.3mm}
\put(50,70){\line(1,0){50}}
\linethickness{0.3mm}
\multiput(80,50)(0.12,0.12){167}{\line(1,0){0.12}}
\linethickness{0.3mm}
\multiput(80,0)(0.12,0.12){167}{\line(1,0){0.12}}
\linethickness{0.3mm}
\put(100,20){\line(0,1){50}}
\linethickness{0.1mm}
\put(50,20){\line(0,1){50}}
\linethickness{0.1mm}
\multiput(30,0)(0.12,0.12){167}{\line(1,0){0.12}}
\linethickness{0.1mm}
\put(50,20){\line(1,0){50}}
\put(25,50){\makebox(0,0)[cc]{$Q_1$}}

\put(25,0){\makebox(0,0)[cc]{$P_1$}}

\put(85,50){\makebox(0,0)[cc]{$Q_2$}}

\put(85,0){\makebox(0,0)[cc]{$P_2$}}

\put(55,73){\makebox(0,0)[cc]{$Q_4$}}

\put(55,23){\makebox(0,0)[cc]{$P_4$}}

\put(100,73){\makebox(0,0)[cc]{$-Q_3$}}

\put(94,23){\makebox(0,0)[cc]{$-P_3$}}

\put(100,45){\makebox(0,0)[cc]{$\downarrow$}}

\put(50,45){\makebox(0,0)[cc]{$\downarrow$}}

\put(80,25){\makebox(0,0)[cc]{$\downarrow$}}

\put(30,25){\makebox(0,0)[cc]{$\downarrow$}}

\put(55,-.3){\makebox(0,0)[cc]{$\rightarrow$}}

\put(55,49.7){\makebox(0,0)[cc]{$\rightarrow$}}

\put(75,69.7){\makebox(0,0)[cc]{$\rightarrow$}}

\put(75,19.7){\makebox(0,0)[cc]{$\rightarrow$}}

\put(40,60){\makebox(0,0)[cc]{$\nearrow$}}

\put(90,60){\makebox(0,0)[cc]{$\nearrow$}}

\put(40,10){\makebox(0,0)[cc]{$\nearrow$}}

\put(90,10){\makebox(0,0)[cc]{$\nearrow$}}

\end{picture}

}

\def\figonesmall{

\def\JPicScale{0.4}
\ifx\JPicScale\undefined\def\JPicScale{1}\fi
\unitlength \JPicScale mm
\begin{picture}(105,75)(0,0)
\linethickness{0.3mm}
\put(30,50){\line(1,0){50}}
\linethickness{0.3mm}
\put(30,0){\line(0,1){50}}
\linethickness{0.3mm}
\put(30,0){\line(1,0){50}}
\linethickness{0.3mm}
\put(80,0){\line(0,1){50}}
\linethickness{0.3mm}
\multiput(30,50)(0.12,0.12){167}{\line(1,0){0.12}}
\linethickness{0.3mm}
\put(50,70){\line(1,0){50}}
\linethickness{0.3mm}
\multiput(80,50)(0.12,0.12){167}{\line(1,0){0.12}}
\linethickness{0.3mm}
\multiput(80,0)(0.12,0.12){167}{\line(1,0){0.12}}
\linethickness{0.3mm}
\put(100,20){\line(0,1){50}}
\linethickness{0.1mm}
\put(50,20){\line(0,1){50}}
\linethickness{0.1mm}
\multiput(30,0)(0.12,0.12){167}{\line(1,0){0.12}}
\linethickness{0.1mm}
\put(50,20){\line(1,0){50}}
\put(24,50){\makebox(0,0)[cc]{{\tiny $Q_1$}}}

\put(24,0){\makebox(0,0)[cc]{{\tiny $P_1$}}}

\put(87,48){\makebox(0,0)[cc]{{\tiny $Q_2$}}}

\put(87,0){\makebox(0,0)[cc]{{\tiny $P_2$}}}

\put(55,75){\makebox(0,0)[cc]{{\tiny $Q_4$}}}

\put(55,25){\makebox(0,0)[cc]{{\tiny $P_4$}}}

\put(100,75){\makebox(0,0)[cc]{{\tiny $-Q_3$}}}

\put(92,25){\makebox(0,0)[cc]{{\tiny $-P_3$}}}

\end{picture}

}

\def\figonesmall{

\def\JPicScale{0.6}
\ifx\JPicScale\undefined\def\JPicScale{1}\fi
\unitlength \JPicScale mm
\begin{picture}(105,75)(0,0)
\linethickness{0.3mm}
\put(30,50){\line(1,0){50}}
\linethickness{0.3mm}
\put(30,0){\line(0,1){50}}
\linethickness{0.3mm}
\put(30,0){\line(1,0){50}}
\linethickness{0.3mm}
\put(80,0){\line(0,1){50}}
\linethickness{0.3mm}
\multiput(30,50)(0.12,0.12){167}{\line(1,0){0.12}}
\linethickness{0.3mm}
\put(50,70){\line(1,0){50}}
\linethickness{0.3mm}
\multiput(80,50)(0.12,0.12){167}{\line(1,0){0.12}}
\linethickness{0.3mm}
\multiput(80,0)(0.12,0.12){167}{\line(1,0){0.12}}
\linethickness{0.3mm}
\put(100,20){\line(0,1){50}}
\linethickness{0.1mm}
\put(50,20){\line(0,1){50}}
\linethickness{0.1mm}
\multiput(30,0)(0.12,0.12){167}{\line(1,0){0.12}}
\linethickness{0.1mm}
\put(50,20){\line(1,0){50}}
\put(26,50){\makebox(0,0)[cc]{{\tiny $Q_1$}}}

\put(26,0){\makebox(0,0)[cc]{{\tiny $P_1$}}}

\put(85,50){\makebox(0,0)[cc]{{\tiny $Q_2$}}}

\put(85,0){\makebox(0,0)[cc]{{\tiny $P_2$}}}

\put(53,73){\makebox(0,0)[cc]{{\tiny $Q_4$}}}

\put(54,23){\makebox(0,0)[cc]{{\tiny $P_4$}}}

\put(100,73){\makebox(0,0)[cc]{{\tiny $-Q_3$}}}

\put(95,23){\makebox(0,0)[cc]{{\tiny $-P_3$}}}

\end{picture}

}

\def\figonealt{
 
\def\JPicScale{0.8}
\ifx\JPicScale\undefined\def\JPicScale{1}\fi
\unitlength \JPicScale mm
\begin{picture}(105,75)(0,0)
\linethickness{0.3mm}
\put(30,50){\line(1,0){50}}
\linethickness{0.3mm}
\put(30,0){\line(0,1){50}}
\linethickness{0.3mm}
\put(30,0){\line(1,0){50}}
\linethickness{0.3mm}
\put(80,0){\line(0,1){50}}
\linethickness{0.3mm}
\multiput(30,50)(0.12,0.12){167}{\line(1,0){0.12}}
\linethickness{0.3mm}
\put(50,70){\line(1,0){50}}
\linethickness{0.3mm}
\multiput(80,50)(0.12,0.12){167}{\line(1,0){0.12}}
\linethickness{0.3mm}
\multiput(80,0)(0.12,0.12){167}{\line(1,0){0.12}}
\linethickness{0.3mm}
\put(100,20){\line(0,1){50}}
\linethickness{0.1mm}
\put(50,20){\line(0,1){50}}
\linethickness{0.1mm}
\multiput(30,0)(0.12,0.12){167}{\line(1,0){0.12}}
\linethickness{0.1mm}
\put(50,20){\line(1,0){50}}
\put(25,50){\makebox(0,0)[cc]{$a$}}

\put(25,0){\makebox(0,0)[cc]{$e$}}

\put(85,50){\makebox(0,0)[cc]{$b$}}

\put(85,0){\makebox(0,0)[cc]{$f$}}

\put(55,73){\makebox(0,0)[cc]{$0$}}

\put(55,23){\makebox(0,0)[cc]{$-a_2$}}

\put(100,73){\makebox(0,0)[cc]{$-a_1$}}

\put(94,23){\makebox(0,0)[cc]{$h$}}

\end{picture}

}
\def\figtwo{

\def\JPicScale{0.8}
\ifx\JPicScale\undefined\def\JPicScale{1}\fi
\unitlength \JPicScale mm
\begin{picture}(105,75)(0,0)
\linethickness{0.3mm}
\put(30,50){\line(1,0){50}}
\linethickness{0.3mm}
\put(30,0){\line(0,1){50}}
\linethickness{0.3mm}
\put(30,0){\line(1,0){50}}
\linethickness{0.3mm}
\put(80,0){\line(0,1){50}}
\linethickness{0.3mm}
\multiput(30,50)(0.12,0.12){167}{\line(1,0){0.12}}
\linethickness{0.3mm}
\put(50,70){\line(1,0){50}}
\linethickness{0.3mm}
\multiput(80,50)(0.12,0.12){167}{\line(1,0){0.12}}
\linethickness{0.3mm}
\multiput(80,0)(0.12,0.12){167}{\line(1,0){0.12}}
\linethickness{0.3mm}
\put(100,20){\line(0,1){50}}
\linethickness{0.1mm}
\put(50,20){\line(0,1){50}}
\linethickness{0.1mm}
\multiput(30,0)(0.12,0.12){167}{\line(1,0){0.12}}
\linethickness{0.1mm}
\put(50,20){\line(1,0){50}}
\put(25,50){\makebox(0,0)[cc]{$0$}}

\put(25,0){\makebox(0,0)[cc]{$1$}}

\put(85,50){\makebox(0,0)[cc]{$\widehat{Q}_2$}}

\put(85,0){\makebox(0,0)[cc]{$0$}}

\put(55,73){\makebox(0,0)[cc]{$\widehat{Q}_4$}}

\put(55,23){\makebox(0,0)[cc]{$0$}}

\put(100,73){\makebox(0,0)[cc]{$-\widehat{Q}_3$}}

\put(94,23){\makebox(0,0)[cc]{$-\widehat{P}_3$}}

\end{picture}

}

\begin{document}

\parindent=12pt

\baselineskip 24pt

\begin{center}

{\Large \bf Bhargava's Cube and Black Hole Charges}


\end{center}

\baselineskip=18pt

\bigskip

\centerline{Nabamita Banerjee$^a$, Ajit Bhand$^a$, Suvankar Dutta$^a$, 
Ashoke Sen$^b$, Ranveer 
Kumar Singh$^a$}

\bigskip

\centerline{\large \it ~$^a$Indian Institute of Science Education and Research
Bhopal}
\centerline{\large \it Bhopal bypass, Bhopal 462066, India}

\centerline{\large \it ~$^b$Harish-Chandra Research Institute, HBNI}
\centerline{\large \it  Chhatnag Road, Jhusi,
Allahabad 211019, India}

\bigskip

\centerline{E-mail: nabamita,abhand,suvankar,ranveer@iiserb.ac.in, sen@hri.res.in}

\vskip .6cm
\medskip

\vspace*{4.0ex}

\centerline{\bf Abstract} \bigskip

\noindent Black holes in a class of string compactifications, known as STU models, 
carry four electric and four magnetic charges. Furthermore
a duality group, given by the product of three congruence subgroups of 
SL(2,${\mathbb Z}$),
acts on these integer valued charges. By placing these eight charges at
the eight corners of a Bhargava cube, we provide a classification of the duality
orbits in these theories.

\bigskip

\bigskip

\bigskip

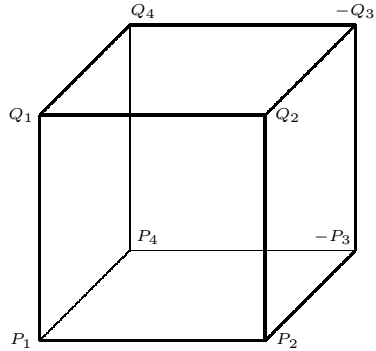
\centerline{\figonesmall}

\vfill\eject

\baselineskip=18pt

\tableofcontents

\sectiono{Introduction} \label{s0}

The relation between black holes in string theory and class groups was first 
explored in \cite{9807056,9807087,0401049}. 
According to this, there is a one to one correspondence between
the duality orbits of supersymmetric 
black holes in heterotic string theory compactified
on a six dimensional torus $T^6$ and the equivalence classes
of positive definite 
binary quadratic forms --
quadratic forms of the form $A x^2 +B xy + C y^2$, 
up to equivalence under $SL(2,\ZZZ)$ transformations on $(x,y)$. 
For a given black hole, the constants $A$, $B$
and $C$ themselves are constructed from T-duality invariant quadratic 
combinations of
electric and magnetic charges.\footnote{This relation holds only for a special class
of charges for which the discriminant $D=B^2-4AC$ is either odd and square free,
or even and $D/4$ is square free and is congruent to 2 or 3 modulo 4. 
For more general charges the
duality orbits require extra data\cite{askitas,0702150,0801.0149,0712.0043}. 
\label{f1}}  
This relation has been explored in more detail
in recent papers\cite{1807.00797,1903.02323}. 
In particular, \cite{1807.00797} listed several open questions regarding this
relation. One of the most interesting of these questions is as follows. 
We call a binary 
quadratic form $Ax^2+Bxy+Cy^2$  primitive if $\gcd(A,B,C)=1$.
It is well known that
the set of equivalence classes of positive definite 
primitive binary quadratic forms with a fixed discriminant
$D=B^2-4AC$ has the structure of an abelian group known as the {\em class group}
 -- given two such equivalence classes,
we have a composition rule that produces a third equivalence class of
binary quadratic forms with the same
discriminant (see {\it e.g.} \cite{lemmermeyer,trifkovic}). 
This would suggest that a pair of supersymmetric black holes, with the 
same discriminant, can be composed to produce a third 
supersymmetric black hole. The question is:
what is the physical interpretation of such compositions?

We shall not try to answer this question. Instead we shall describe a different application
of the composition of binary quadratic forms using the description of 
Bhargava\cite{bhargava}.
In this description, which will be described in detail in \S\ref{s1}, one describes the
composition with the help of a cube, with integers placed at different corners of the
cube. Bhargava's prescription associates a binary quadratic form with every pair of
opposite faces of the cube. Since there are three such pairs, we have three such 
quadratic forms. It can be shown that they all have the same discriminant, and
furthermore, that if we apply class group composition on 
the equivalence classes of these three quadratic 
forms, the result is the identity element of the group. Therefore Bhargava's prescription
gives a new way of describing group composition, by identifying the quadratic form
associated with one pair of faces as the inverse of the composition of the 
quadratic forms associated with the other two pairs of faces.

We shall use Bhargava's construction to address a different problem involving
black hole charges in string theory.\footnote{In string theory one does not always 
distinguish between black holes and other single particle states; our use of the
word black hole should be taken in this spirit.}
In a class of string compactifications known as the
STU models\cite{9508064,9608059,9901117,0512227,0702187,1907.04077}, 
any state is characterized by a 4 dimensional 
electric charge vector $Q$ and a 4 dimensional magnetic charge vector $P$, normalized
to have integer entries. 
They are
acted upon by the T-duality group $\Gamma_U\times\Gamma_T$
and the S-duality group
$\Gamma_S$, where $\Gamma_U$, $\Gamma_T$ and $\Gamma_S$ are
isomorphic to some congruence subgroups of $SL(2,\ZZZ )$.
The problem we want to address is:  
given two charge vectors $(Q,P)$ and $(Q',P')$, 
how do we know if they are in the same duality orbit, i.e.\ that they
can be related by a 
$\Gamma_S\times\Gamma_T\times\Gamma_U$ transformation? 
Unlike in the case of heterotic string theory on
$T^6$, where for special class of charges discussed in footnote \ref{f1}, the
orbit is classified by the equivalence class of a single quadratic form composed from
T-duality invariant quadratic combination of the charges, the answer here is more
complicated.
We shall show that a compact answer to this question is provided by the Bhargava 
cube\cite{bhargava}, 
with the eight components of the electric and magnetic charges sitting at the
eight corners of the cube. This also provides physical interpretation of the integers
sitting at the corners of the Bhargava cube -- they are the charges carried by the
black holes of the STU model. In other words, each Bhargava cube describes a single black hole in the STU model, unlike in \cite{1807.00797,1903.02323}, where it
would correspond to a map 
from a pair of 
black holes to a third black hole.

We shall restrict our analysis to
charges for which the discriminant $B^2-4AC$
of the associated quadratic forms is negative, since
only in this case a single black hole carrying these charges has regular event horizon.
This requires $AC>0$. However, we shall not assume that $A$ and $C$ themselves
are positive. Therefore the quadratic forms can be either positive definite or negative
definite. We shall also not assume that the quadratic forms are primitive. Furthermore, 
we shall 
assume that the charges can take arbitrary integer values. In specific models,
there may be further
duality invariant constraints on the charges that depend on the details of the model. 
These may involve imposing positive definiteness on some of the quadratic forms,
and / or restricting the charges to belong to a sublattice of the integer lattice.
This will eliminate some of the duality orbits that we find in our analysis.

The problem of classification of the duality orbits under 
$\Gamma_S\times\Gamma_T
\times \Gamma_U$ could also arise in more general situations, {\it e.g.} in the
case of heterotic or type II string theories compactified on six dimensional torus
$T^6$. In these cases the duality groups are much larger, {\it e.g.}
$O(6,22;\ZZZ )\times SL(2,\ZZZ )$ for heterotic on $T^6$ and $E_{7,7}(\ZZZ )$ for type II
on $T^6$. The duality transformations act simultaneously on the charges and the
moduli space of the theory. However we may choose to work in
a subspace of the full moduli space of the theory
which is invariant under some discrete 
symmetry group $G$, and examine the spectrum of states carrying definite representation
of $G$, or consider twisted index where we count states weighted by 
the elements of $G$\cite{0911.1563}. 
In this case only the part of the duality group that commutes
with $G$ will be a symmetry of the $G$-twisted index.  
This part of the duality
group can be much smaller and in special cases may have the structure of a product
of congruence subgroups of $SL(2,\ZZZ )$. 
Furthermore, questions of this type are well defined only for states carrying
charge vectors that are
invariant under $G$, and in special cases these could be four dimensional electric and
magnetic charges on which the residual duality group acts.
A concrete example of this  
may be provided as follows. One possible stringy
embedding of the STU model (given in example D in \cite{9508064}) 
is to take type II string
theory on $T^4\times T^2$, and take the quotient of the theory by a 
$\ZZZ _2\times \ZZZ _2$
discrete symmetry $G$
associated with $T^4$ compactification, accompanied by a translation along $T^2$.
$G$ commutes with only an $SL(2,\ZZZ )^3$ subgroup
of the duality group, and the translation along $T^2$ commutes with only a
$\Gamma_0(2)^3$ subgroup of the $SL(2,\ZZZ )^3$, breaking the duality symmetry to
$\Gamma_0(2)^3$. However, 
instead of taking the quotient by the discrete symmetry, if we
consider the $G$ twisted index in type II on $T^4\times T^2$\cite{0911.1563}, 
then the
relevant symmetry of the twisted index is $SL(2,\ZZZ )^3$, since in the absence of
translation along $T^2$ there is no breaking of $SL(2,\ZZZ )^3$ to $\Gamma_0(2)^3$.
The charge vectors invariant under $G$ are four component electric and four component
magnetic charges, on which the residual $SL(2,\ZZZ )^3$ duality group acts.
Therefore in this situation we would require classification of the duality orbits under the
$SL(2,\ZZZ )_S\times SL(2,\ZZZ )_T\times SL(2,\ZZZ )_U$ symmetry. Similar
analysis can be carried out for other string compactifications whose exact duality
group contains as a subgroup product of three congruence subgroups of $SL(2,\ZZZ)$,
{\it e.g.} for the FHSV model\cite{9505162}.

We now give a short summary of our results. 
In \S\ref{s1} we study the duality orbits
of $SL(2,\ZZZ )_S\times SL(2,\ZZZ )_T\times SL(2,\ZZZ )_U$. With 
the eight component
charge vector $(Q_1,\ldots, Q_4, P_1,\ldots P_4)$ we can construct three quadratic forms,
each being invariant under two of the $SL(2,\ZZZ)$ groups and transforming under
the third one. Therefore we have
three equivalence classes. However by placing the eight components of charges at
the eight corners of a Bhargava cube, we show that when the quadratic forms are primitive,
i.e.\ the coefficients $A$, $B$ and $C$ appearing in each of the quadratic forms are
coprime,\footnote{If two of the quadratic forms are primitive, then one can show that the
third one is automatically primitive.} 
then the equivalence class of the third quadratic form is uniquely determined
in terms of the equivalence classes of the other two. 
In fact, we shall show that the full charge vector $(Q_1,\ldots, Q_4, P_1,\ldots P_4)$ is
determined, up to duality transformation, by the equivalence classes of two of the
quadratic forms.
Therefore the duality orbits are
completely classified by equivalence classes of a pair of quadratic forms. In \S\ref{s2}
we generalize the analysis to the case where the quadratic forms are not primitive.
We call a pair of quadratic forms coprime if they do not have a common
integer factor.
We show that if two of the quadratic forms are coprime, even though they may not
be individually primitive, the charge vector is still uniquely 
determined, up to duality transformation, 
by the equivalence classes of the first two quadratic forms. Therefore
the duality orbits continue to be classified by the equivalence classes of a pair of
quadratic forms. When the two quadratic forms are not coprime, but have a
common integer factor $m>1$, then the duality orbit is classified by the equivalence 
classes of the two quadratic forms, the integer $m$, 
and a pair of integers
$(d,b)$, where $d$ is a factor of $m$ and $0\le b\le (d-1)$.
However, special care is needed if any of the quadratic forms is
proportional to either $x^2+y^2$ or $x^2+xy+y^2$ up to $SL(2,\ZZZ)$ transformation,
since these have non-trivial 
automorphisms -- $SL(2,\ZZZ)$ transformations under which these
quadratic forms remain invariant\cite{buell,vol}. In this case 
there are additional identifications between the duality orbits labelled by 
$(a\equiv m/d,b,d)$. 
In particular if any of the quadratic forms is proportional to $x^2+y^2$,
then $(a,b,d)$ and $(a',b',d')$ describe the same duality orbit if $a'=\gcd(b,d)$,
$d'=ad/\gcd(b,d)$
and $b'$ is related to $b$ in a more complicated fashion explained in
\refb{efin11}, \refb{efin12}. 
On the other hand if any of the quadratic forms is proportional
to $x^2+xy+y^2$, then $(a,b,d)$ and $(a',b',d')$ describe the same
duality orbit provided $a'=\gcd(b,d)$, $d'=ad/\gcd(b,d)$,
and $b'$ 
is determined by solving \refb{3aa}. In this case there is a further identification -- 
the orbits labelled by $(a,b,d)$ and $(a',b',d')$  are also the same
if
$a'=\gcd(d,a+b)$, $d'=ad/\gcd(d,a+b)$, and 
$b'$ is obtained by solving \refb{e4de}.

In \S\ref{s3} we turn to the problem of classification of the duality orbits of 
$\Gamma_S\times\Gamma_T\times\Gamma_U$, where $\Gamma_S$,
$\Gamma_T$ and $\Gamma_U$ are congruence
subgroups of $SL(2,\ZZZ)$, but are not necessarily the same congruence subgroups.
In this case we need the notion of $\Gamma$ equivalence classes\cite{1711.00230} 
-- two quadratic
forms are considered equivalent if they are related by a $\Gamma$-transformation,
$\Gamma$ being a congruence subgroup of $SL(2,\ZZZ)$. As before,
we need to pick two of the three quadratic forms associated with the charge
vector -- for definiteness we take them
to be those that transform under $\Gamma_S$ and $\Gamma_U$. Therefore they are
classified by $\Gamma_S$ and $\Gamma_U$ equivalence classes respectively. We show
that if these quadratic forms are coprime, then the duality orbit is classified by their
$\Gamma_S$ and $\Gamma_U$ equivalence classes, 
and the choice of an element of
the coset $\Gamma_T\backslash 
SL(2,\ZZZ)$. On the other hand if the two quadratic forms 
have a common factor 
$m$, then besides the data described above, and $m$,
we also need to choose a pair of integers
$(d,b)$, where $d$ is a factor of $m$ and $0\le b\le (d-1)$.
Finally for special quadratic forms, related to $(x^2+y^2)$ or $(x^2+xy+y^2)$
by $SL(2,\ZZZ)$ transformation up to a multiplicative constant, some of the
duality orbits labelled by $m$, $b$, $d$ and the elements of 
$\Gamma_T\backslash SL(2,Z)$, get identified. The procedure for determining when
that happens has been explained in \S\ref{s3}.

In Appendix \ref{sa} we review the proof of Gauss's
Lemma dealing with properties of $2\times n$ matrices with integer entries. 
This is 
well known to 
mathematicians but may
not be so well-known to physicists. 
In Appendix \ref{sb} we describe a proof of the existence of Bhargava's cube for a 
given pair of binary quadratic forms even when they are not primitive or
coprime, generalizing the corresponding result 
for primitive
quadratic forms. For a pair of
primitive quadratic forms, the Bhargava cube is known to be 
unique up to $SL(2,\ZZZ)$ transformations, but this is not so when 
the two quadratic forms are not primitive and not coprime.
Appendix \ref{sbhargava} contains a brief review of the relevant parts of 
Bhargava's
results that we use in our analysis.
Appendix \ref{smiddle} reviews properties of some special quadratic forms
that remain invariant under some $SL(2,\ZZZ)$ transformations, since our analysis
runs into additional subtleties for such special quadratic forms.
Appendix \ref{sc} contains a technical result that is necessary to classify the
duality orbits when we have such special quadratic forms. In 
Appendix \ref{sd} we count the number of duality
orbits in several examples, verifying that the number is invariant under
permutations of $\Gamma_S$, $\Gamma_T$ and $\Gamma_U$, even though the
algorithm we give treats $\Gamma_T$ differently from 
$\Gamma_S$ and $\Gamma_U$.

Note added: After submitting the paper to the arXiv, we became aware of 
Refs.~\cite{borsten1,borsten2} where possible connection between Bhargava's
cube and the black hole charges in the STU model was discussed. Although the
original references are not available in the arXiv, a review of these results can be
found in the recent arXiv paper \cite{2006.03574}.

\sectiono{$SL(2,\ZZZ )^3$ orbits: Primitive case} \label{s1}

In  this section we shall consider the case where
the duality group is
$SL(2,\ZZZ )_S\times SL(2,\ZZZ )_T\times SL(2,\ZZZ )_U$.
We denote the four components of the electric charge vector $Q$ by $Q_1,\ldots,Q_4$
and the four components of the magnetic charge vector $P$ by $P_1,\ldots, P_4$.
The charges are normalized so that $Q_i,P_i\in \ZZZ$ for $1\le i\le 4$.
Duality transformation by an element of 
$SL(2,\ZZZ )_S\times SL(2,\ZZZ )_T\times SL(2,\ZZZ )_U$ 
gives a map 
$(Q_1,\ldots, Q_4,P_1,\ldots, P_4)\in\ZZZ^8\mapsto (Q_1',\ldots, Q_4',P_1',\ldots, P_4')
\in\ZZZ^8$,
defined as follows.
$SL(2,\ZZZ )_U$ acts by left multiplication on the vectors:
\be\label{e2}
\begin{pmatrix} Q_1\cr Q_2 \end{pmatrix}, \, \begin{pmatrix} Q_4\cr -Q_3\end{pmatrix}, 
\, \begin{pmatrix} P_1\cr P_2\end{pmatrix}, \, 
\begin{pmatrix} P_4\cr -P_3\end{pmatrix}\, .
\ee
This means, for example, that under 
an $SL(2,\ZZZ)_U$ transformation by a matrix $W$, we have
$\begin{pmatrix} Q_1'\cr Q_2' \end{pmatrix}= W\begin{pmatrix} Q_1\cr Q_2 \end{pmatrix}$.
Similarly, 
$SL(2,\ZZZ )_T$ acts by left multiplication on the vectors:
\be\label{e3}
\begin{pmatrix} Q_1\cr Q_4 \end{pmatrix}, \, \begin{pmatrix} Q_2\cr -Q_3
\end{pmatrix}, \, \begin{pmatrix} P_1\cr P_4\end{pmatrix}, \, 
\begin{pmatrix} P_2\cr -P_3\end{pmatrix}\, .
\ee
Finally $SL(2,\ZZZ )_S$ acts by left multiplication on the vectors:
\be\label{e4}
\begin{pmatrix} Q_1\cr P_1\end{pmatrix}, \, \begin{pmatrix} Q_2\cr P_2
\end{pmatrix}, \, \begin{pmatrix} Q_3\cr P_3\end{pmatrix}, \, 
\begin{pmatrix} Q_4\cr P_4\end{pmatrix}\, .
\ee
In string theory, these transformation laws follow from the interpretation of 
$Q_1$ and $Q_2$ as the
momentum quantum numbers along two circles that form part of the
compact internal space, $Q_3$ and $Q_4$ as the winding
numbers along the same two circles and $P_1,\ldots, P_4$ as the magnetic 
counterparts of these charges.

The elements of the $\ZZZ^8$ charge lattice can be grouped into equivalence 
classes -- two elements are considered equivalent if there exists a duality
transformation that maps one element to the other. We shall call these 
equivalence classes
duality orbits. Our goal will be to classify the duality orbits  of the theory.

Now given a pair of vectors $\begin{pmatrix} w_1\cr z_1\end{pmatrix}$ and 
$\begin{pmatrix} w_2
\cr z_2\end{pmatrix}$ on which an $SL(2,\ZZZ)$ transformation acts by left multiplication,
the combination
$(w_1z_2-z_1w_2)$ 
is $SL(2,\ZZZ )$ invariant. Using this we get the following quadratic invariants
of $SL(2,\ZZZ )_U$, $SL(2,\ZZZ )_T$ and $SL(2,\ZZZ )_S$:
\ben 
SL(2,\ZZZ )_U &:& 
Q_1 Q_3+Q_2Q_4, \quad P_1 P_3+P_2P_4, \quad Q_1P_2-Q_2P_1,
\nonumber \\ &&
Q_3P_4-Q_4P_3,\quad Q_1P_3+Q_2P_4, \quad Q_4P_2+Q_3P_1\, , \nonumber \\
SL(2,\ZZZ )_T &:& Q_1 Q_3+Q_2Q_4, \quad P_1 P_3+P_2P_4, \quad Q_1P_4-Q_4P_1,
\nonumber \\ &&
Q_2P_3-Q_3P_2,\quad Q_1P_3+Q_4P_2, \quad Q_2P_4+Q_3P_1\, , \nonumber \\
SL(2,\ZZZ )_S &:& Q_1P_2-Q_2P_1, \quad 
Q_1 P_3-Q_3 P_1, \quad  Q_1P_4-Q_4P_1,
\nonumber \\ &&
Q_2P_3-Q_3P_2,\quad Q_2P_4-Q_4P_2, \quad Q_3P_4-Q_4P_3\, .
 \een

\begin{figure}
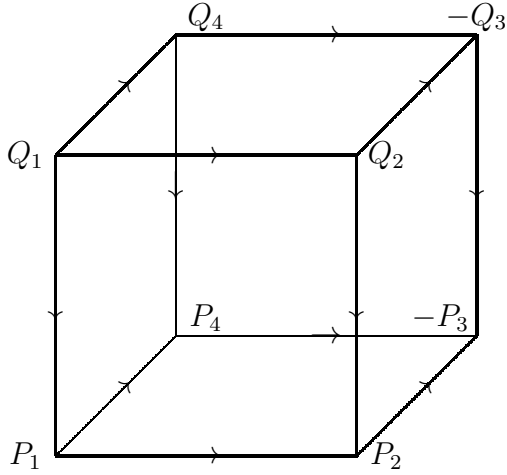


\begin{center}

\figone

\end{center}

\bigskip

\caption{The Bhargava cube labelled by the eight dimensional dyonic charge.
$\downarrow$ pairs charges that transform under $SL(2,\ZZZ)_S$, $\rightarrow$
pairs charges that transform under $SL(2,\ZZZ)_U$ and $\nearrow$ pairs charges that
transform under $SL(2,\ZZZ)_T$.
\label{figone}}

\end{figure}

Consider now a Bhargava cube with the entries at the eight corners as shown in
Fig.~\ref{figone}. 
The cube has the property that a pair of entries connected by
a top$\to$down line transforms under $SL(2,\ZZZ )_S$, 
a 
pair of entries connected by a left$\to$right line transforms under 
$SL(2,\ZZZ )_U$ and a 
pair of entries connected by a front$\to$back line transforms under 
$SL(2,\ZZZ )_T$. 
We can now construct three binary quadratic 
forms associated with the three opposite pairs of faces of 
this cube. 
For a given pair of faces, the quadratic form is given by $-\det(Mx+Ny)$, where $M$
and $N$ are the integer 
matrices associated with the first and the second faces respectively.
The top left corner of $M$ is always chosen to be the same integer, which 
we shall take to be $Q_1$. With this rule,
the quadratic form
associated with the top-down faces is
\ben\label{ex1}
q_1&=& -\det \begin{pmatrix} Q_1 x + P_1 y & Q_2 x + P_2 y\cr Q_4 x + P_4 y & -Q_3 x - P_3 y 
\end{pmatrix}
\nonumber \\
&=& (Q_1 Q_3 + Q_2 Q_4) x^2 + (P_1 P_3+P_2P_4)y^2 + (Q_1 P_3+Q_3 P_1
+Q_2 P_4 + Q_4 P_2) xy\, .
\een
The coefficients are $SL(2,\ZZZ )_U\times SL(2,\ZZZ )_T$ invariant combinations.
Similarly the quadratic form
associated with the left-right faces takes the form:
\ben\label{ex3}
q_2&=& -\det \begin{pmatrix} Q_1 x + Q_2 y & P_1 x + P_2 y \cr Q_4 x - Q_3 y & P_4 x - P_3 y 
\end{pmatrix}
\nonumber \\
&=& (Q_4 P_1-Q_1 P_4) x^2 + (Q_2 P_3-Q_3P_2)y^2 + (Q_4 P_2-Q_2 P_4
-Q_3 P_1 + Q_1 P_3) xy\, ,
\een
which has $SL(2,\ZZZ )_S\times SL(2,\ZZZ )_T$ invariant coefficients. Finally, 
the quadratic form associated with the front-back faces is given by
\ben\label{ex2}
q_3&=& -\det \begin{pmatrix} Q_1 x + Q_4 y & Q_2 x - Q_3 y \cr P_1 x + P_4 y & P_2 x - P_3 y
\end{pmatrix}
\nonumber \\
&=& (Q_2 P_1-Q_1 P_2) x^2 + (Q_4 P_3-Q_3P_4)y^2 + (Q_1 P_3-Q_3 P_1
-Q_4 P_2 + Q_2 P_4) xy\, ,
\een
with $SL(2,\ZZZ )_S\times SL(2,\ZZZ )_U$ 
invariant coefficients.

Various duality transformations described in \refb{e2}-\refb{e4} 
by the $SL(2,\ZZZ)$ matrix 
$\begin{pmatrix} \alpha & \beta\cr\gamma &\delta \end{pmatrix}$ can be regarded as
transformations of the variables $(x,y)$ of the form:
\be\label{etrsxy}
(x,y)\mapsto (x',y')=(x,y) \begin{pmatrix} \alpha & \beta\cr\gamma &\delta \end{pmatrix}\, .
\ee
For example the $SL(2,\ZZZ )_S$ transformation on $Q_i$'s and $P_i$'s described in
\refb{e4} by the matrix 
$\begin{pmatrix} \alpha & \beta\cr\gamma &\delta \end{pmatrix}$
transforms the quadratic form \refb{ex1} to a new one 
that can be obtained by
making the replacement \refb{etrsxy} in the original quadratic form. Similar remark holds
for $SL(2,\ZZZ )_U$ and $SL(2,\ZZZ )_T$ transformations.

Given a quadratic form $Ax^2+Bxy+Cy^2$, we 
have defined the discriminant $D$ to be
$B^2-4AC$. It is easy to verify that 
all of the quadratic forms \refb{ex1}, \refb{ex3} and \refb{ex2} 
have the same discriminant $D$ given by the
$SL(2,\ZZZ )_S\times SL(2,\ZZZ )_T\times SL(2,\ZZZ )_U$  invariant
quartic combination of the charges:
\be
D=-4\, (Q_1 Q_3 + Q_2 Q_4)\, (P_1 P_3+P_2P_4)+(Q_1 P_3+Q_3 P_1
+Q_2 P_4 + Q_4 P_2)^2\, .
\ee

We now turn to the problem of classification of the duality orbits of 
$SL(2,\ZZZ )_S\times SL(2,\ZZZ )_T\times SL(2,\ZZZ )_U$. In this section, we
shall consider the case where
each of the
quadratic forms given in \refb{ex1}, \refb{ex3} and \refb{ex2} is
primitive, i.e.\ $A,B,C$ are coprime for each of them. 
It can be easily argued that this condition is invariant under the duality group, i.e.\ 
starting with a cube where all three quadratic forms are primitive, we cannot make a 
duality transformation that makes one or more of them non-primitive. In this case,
to each of the quadratic forms we can associate an element of the class group for
discriminant $D$, consisting of the equivalence classes of binary quadratic forms up to
$SL(2,\ZZZ )$ transformations. Naively, there will be three such elements 
that will characterize a  given duality orbit. However, since the three class group
elements are
associated with three opposite pairs of faces of a Bhargava cube, 
the composition of the three elements gives the 
identity element of the
group by Bhargava's theorem\cite{bhargava}.
Therefore only two of them are independent.
It also follows from Bhargava's result that two different charge vectors,
for which the corresponding Bhargava cubes have the same (primitive)
class group elements, can be
related to each other by duality transformation. 
The proof of these results have been reviewed in Appendix \ref{sbhargava}.
Therefore we can use any two of
the three class group elements associated with the quadratic forms 
\refb{ex1}, \refb{ex3} and \refb{ex2} to characterize a duality orbit of 
$SL(2,\ZZZ )_S\times SL(2,\ZZZ )_T\times SL(2,\ZZZ )_U$. 
For definiteness we shall choose the
class group elements associated with the quadratic forms \refb{ex1} and \refb{ex3} to
label the duality orbit. As is well known\cite{cohen,cox,trifkovic}, 
these class group elements can be
represented uniquely by quadratic forms of the type 
$Ax^2+Bxy+Cy^2$ with $A,B,C$ satisfying the following conditions:
\be \label{erange1}
|B|\le |A|\le |C|, \quad AB>0 \quad \hbox{for} \quad 
|B|=|A| \quad \hbox{or} \quad A=C\, .
\ee

\sectiono{$SL(2,\ZZZ)^3$ orbits: Non-primitive case} \label{s2}

Only primitive quadratic forms $Ax^2+Bxy+Cy^2$  form elements
of class groups obeying group composition laws. However even when the quadratic
form is not primitive, one can still define the notion of equivalence classes under 
$SL(2,\ZZZ )$ transformations, {\it e.g.} different quadratic forms \refb{ex1} related 
by $SL(2,\ZZZ )_S$ transformations are considered equivalent. Furthermore, the inequivalent
quadratic forms are still characterized by restricting the coefficients to satisfy 
\refb{erange1}, but now we do not demand $A$, $B$ and $C$ to be coprime. 
Therefore a natural question
would be: even when the quadratic forms \refb{ex1} and
/ or \refb{ex3} are not primitive, do their equivalence classes under $SL(2,\ZZZ )_S$ and
$SL(2,\ZZZ )_U$ transformations determine the duality orbit uniquely?

We begin by introducing some notations. Let us  denote the quadratic forms \refb{ex1}
and \refb{ex3} by:
\ben \label{enon1}
q_1 &\equiv& A_1 x^2 + B_1 xy + C_1 y^2 \nonumber \\ &=&
(Q_1 Q_3 + Q_2 Q_4) \, x^2 + (P_1 P_3+P_2P_4)\, y^2 + (Q_1 P_3+Q_3 P_1
+Q_2 P_4 + Q_4 P_2) \, xy, \nonumber \\
q_2 &\equiv& A_2 x^2 + B_2 xy+ C_2 y^2 \nonumber \\
&=&(Q_4 P_1-Q_1 P_4) \, x^2 + (Q_2 P_3-Q_3P_2)\, y^2 + (Q_4 P_2-Q_2 P_4
-Q_3 P_1 + Q_1 P_3) \, xy
\, .
\een
We also define the minor matrix for the Bhargava cube $\AAA$ depicted in
Fig.~\ref{figone}  as
\begin{equation} \label{enon2}
M(\mathcal{A})=\begin{pmatrix}
Q_1&Q_2&P_1&P_2\\ Q_4&-Q_3&P_4&-P_3
\end{pmatrix}\, .
\end{equation} 
Clearly, the minor matrix carries exactly the same information as the 
Bhargava cube, but the information is represented in an asymmetric manner.
It can be easily checked that 
\begin{equation}
\begin{split}
M_{12} = -A_1 ,~~~~~M_{13} = -A_2 ,~~~~~ M_{14} = -\frac{B_2+B_1}{2},
\\
M_{34} = -C_1 ,~~~~~ M_{24} = -C_2 , ~~~~~M_{23} = -\frac{B_2-B_1}{2},
\end{split}
\label{1}
\end{equation}
where $M_{ij}$ denotes the minor of the matrix $M(\mathcal{A})$, obtained by keeping 
only the $i$'th and $j$'th column of $M(\AAA)$ and then taking the determinant 
of the
corresponding square matrix. 
Note that $(B_1 \pm B_2 )/2$ are integers since $q_1$ and $q_2$ have the same 
discriminants.

Now we record a Lemma due to Gauss which we need to carry out our analysis. 
\begin{lemma} (Gauss's Lemma)
Let 
\begin{equation}
M=\begin{pmatrix}
p_1&p_2&\dots &p_n\\
r_1&r_2 &\dots &r_n
\end{pmatrix}~~~
\text{and}~~~M'=\begin{pmatrix}
p_1'&p_2'&\dots &p_n'\\
r_1'&r_2' &\dots &r_n'
\end{pmatrix}
\end{equation}
be two $2 \times n-$matrices ($n \geq 3$) with the following properties:
\begin{enumerate}
\item the $2 \times 2-$minors of $M$ are coprime;
\item  there is an integer $m$ such that each minor of $M'$ is $m$ times the corresponding minor
of $M$. \end{enumerate}
Then there is a matrix $g =\begin{pmatrix}
a&b\\c&d
\end{pmatrix}$ with determinant $m$ and $a,b,c,d\in\ZZZ$, such that $M'= gM$.

\end{lemma}
The proof of this Lemma has been reviewed in Appendix \ref{sa}.

We now return to our problem. Suppose that 
the quadratic forms $q_1$ and $q_2$ given
in \refb{enon1} are not primitive, but that $(A_1,B_1,C_1,A_2,B_2,C_2)$ are coprime. 
This implies that $(A_1,C_1,A_2,C_2,(B_1+B_2)/2,(B_1-B_2)/2)$ 
are also coprime. 
Now suppose that we have another set of charges $(Q_1',\ldots, Q_4',P_1',\ldots, P_4')$
with the same quadratic forms $q_1$ and $q_2$ given in \refb{enon1}. Then by
Gauss's Lemma the corresponding minor matrix
\begin{equation} \label{enon4}
M(\mathcal{A}')=\begin{pmatrix}
Q_1'&Q_2'&P_1'&P_2'\\ Q_4'&-Q_3'&P_4'&-P_3'
\end{pmatrix}
\end{equation} 
must be related to $M(\AAA)$ defined in \refb{enon2} by an $SL(2,\ZZZ )$ 
transformation:
\be
M(\AAA')=\begin{pmatrix}  \alpha & \beta\cr \gamma & \delta
\end{pmatrix} M(\AAA), \quad
\alpha,\beta,\gamma,\delta\in \ZZZ , \quad \alpha\delta-\beta\gamma=1\, .
\ee
Using \refb{e3} we can identify this as an $SL(2,\ZZZ )_T$ transformation. Therefore the
charge vectors $(Q_1,\ldots, Q_4, P_1,\ldots, P_4)$ and $(Q_1',\ldots, Q_4', P_1',
\ldots,  P_4')$ are in the same duality
orbit.

If we consider the more general case where the quadratic forms $q_1'$ and $q_2'$
associated with the charge vector $(Q_1',\ldots, Q_4', P_1',
\ldots,  P_4')$ are not the same as $q_1$ and
$q_2$ but are in the same equivalence classes, then we can first apply an $SL(2,\ZZZ )_S$
transformation to make $q_1'=q_1$, and then apply an $SL(2,\ZZZ )_U$ transformation to
make $q_2'=q_2$. Then we can use Gauss's Lemma to show that the two charge vectors
are related by $SL(2,\ZZZ )_T$ transformation.

Therefore we conclude that if $(A_1,B_1,C_1,A_2,B_2,C_2)$ are coprime, then
the equivalence classes of the quadratic forms given in \refb{enon1} uniquely determine
the duality orbit. Put another way, two charge vectors, for which the equivalence 
classes of the quadratic forms
given in \refb{enon1} agree, are related by a duality
transformation.

In order to complete the analysis, we need to check that given a pair of quadratic forms of
the form shown in \refb{enon1}, there exists a Bhargava cube that generates these
quadratic forms. This has been proved in Appendix \ref{sb} following the same method
that is used to prove the existence of Bhargava cube for a pair of primitive quadratic 
forms.

Finally, consider the case where $(A_1,B_1,C_1,A_2,B_2,C_2)$ are not coprime but
have some common factor $m>1$: 
\be \label{enon88}
m=\gcd (A_1,B_1,C_1,A_2,B_2,C_2)\, .
\ee
In that case, we can define another pair of
quadratic forms:
\be
\wt A_1 x^2 + \wt B_1 xy + \wt C_1 y^2={1\over m} (A_1 x^2 + B_1 xy + C_1 y^2),
\quad
\wt A_2 x^2 + \wt B_2 xy + \wt C_2 y^2={1\over m} (A_2 x^2 + B_2 xy + C_2 y^2),
\ee
so that $(\wt A_1,\wt B_1,\wt C_1,\wt A_2,\wt B_2,\wt C_2)$ are coprime. Let $M_0$
be a minor matrix that represents this new pair of quadratic forms via the analog
of \refb{1}. Of course, $M_0$ is only defined up to an $SL(2,\ZZZ )_T$ transformation, but
we choose one of these matrices. It now follows from 
Gauss's Lemma that the original minor matrix $M(\AAA)$ defined in \refb{enon2}
is given by
\be\label{enon77}
M(\AAA)= g\, M_0\, ,
\ee
where $g$ is an integer matrix with  determinant $m$.

Now let us consider another set of charges $(Q_1',\ldots, Q_4',P_1',
\ldots, P_4')$ for which the
associated  quadratic forms are the same  
as $(q_1,q_2)$.\footnote{As before, if the quadratic classes associated with the
charges $(Q_1',\ldots, Q_4',P_1',\ldots, P_4')$ 
are not the same as $(q_1,q_2)$ but are in the same
equivalence classes, we can make them same by appropriate 
$SL(2,\ZZZ )_S\times SL(2,\ZZZ )_U$ transformations.} 
Gauss's Lemma tells us that the minor matrix $M(\AAA')$ associated with the new
charges is given by
\be\label{enon99}
M(\AAA')= g'\, M_0\, ,
\ee
where $g'$ is another integer matrix with  determinant $m$.

Now  the two sets of charge vectors are in the same duality orbit if $M(\AAA)$ and 
$M(\AAA')$ are related by multiplication by an $SL(2,\ZZZ )$ transformation from the
left. Therefore we need to determine whether $g$ and $g'$ are related 
by an $SL(2,\ZZZ )$ transformation from the
left. We shall prove below that any $2\times 2$ matrix $g$ 
with integer entries and determinant $m$
can be written as,
\be \label{enon7}
g = S\, U, \quad U=
\begin{pmatrix} a & b\cr 0 & d\end{pmatrix}, \quad S\in SL(2,\ZZZ ), \quad
a\, d=m\, , \quad a,d>0, \quad 0\le b\le d-1\, .
\ee
Furthermore the decomposition is unique, i.e.\ for any given matrix $g$, the
matrices $S$ and $U$ are determined without any ambiguity. 
Therefore the inequivalent duality orbits will be classified by different values of $a,b,d$,
besides the equivalence classes of the quadratic forms given in \refb{ex1} and
\refb{ex3}. The decomposition
\refb{enon7} generalizes a construction of \cite{buell} for prime values of
$m$.

We now give a proof of the decomposition \refb{enon7} and its 
uniqueness. Since \refb{enon7} gives $\det g=m$, it follows from 
\refb{enon88}, that $\det g>1$. 
We will explicitly construct the matrices $U$ and $S$ satisfying \refb{enon7}. Let
\begin{equation}
g=\begin{pmatrix}
r&s\\t&u
\end{pmatrix},~~~~
U=\begin{pmatrix}
a&b\\0&d
\end{pmatrix}~~~~\text{and}~~~~S=\begin{pmatrix}
\alpha&\beta\\ \gamma&\delta
\end{pmatrix}.
\end{equation}
We will find $a,b,d,\alpha,\beta,\gamma$ and $\delta$ 
in terms of the entries of $g$. 
Observe that $U=S^{-1}g$ gives 
\begin{equation}\label{eproduct}
\begin{pmatrix}
a&b\\0&d
\end{pmatrix}=\begin{pmatrix}
\delta&-\beta\\ -\gamma&\alpha
\end{pmatrix}\begin{pmatrix}
r&s\\t&u
\end{pmatrix}=\begin{pmatrix}
\delta r-\beta t&\delta s-\beta u\\ -\gamma r+\alpha t&-\gamma s+\alpha u
\end{pmatrix}.
\end{equation}
Thus we must have $\gamma r=\alpha t$. We choose $\gamma = t/\gcd(r,t)$ and
$\alpha = r/\gcd(r,t)$. Since this gives $\gcd(\alpha,\gamma)=1$, 
by Euclid's algorithm, we can find integers 
$\beta$ and $\delta$ such that 
$\alpha\delta-\beta\gamma=1$. 
Thus $S$ constructed in this manner has determinant 1. The integers
$a$, $b$ and $d$ can now be read out from the right hand side of \refb{eproduct}.
Taking determinant of
both sides of the equation $g=SU$, we get that $m=ad$.  Since $ad$ is positive, we can
make $a$ and $d$ positive by multiplying $U$ and $S$ by $-I_2$ if necessary.
Here $I_2$ denotes the $2\times 2$ identity matrix.
To get $0\leq b\leq d-1$, suppose $b>d$ or $b<0$. 
Then $b=h d+\widetilde{b}$ where $0\leq \widetilde{b}\leq d-1$ and $h\in \ZZZ $, $h\ne 0$. 
We then have 
\begin{equation}
\begin{pmatrix}
a&b\\0&d
\end{pmatrix}=\begin{pmatrix}
1&h\\0&1
\end{pmatrix}\begin{pmatrix}
a&\wt b\\0&d
\end{pmatrix}\, .
\end{equation}
Thus we have 
\begin{equation}
\begin{pmatrix}
r&s\\t&u
\end{pmatrix}=\begin{pmatrix}
\alpha&\beta\\ \gamma&\delta
\end{pmatrix}\begin{pmatrix}
1&h\\0&1
\end{pmatrix}\begin{pmatrix}
a&\widetilde{b}\\0&d
\end{pmatrix}.
\end{equation}
So rename $S=\begin{pmatrix}
\alpha&\beta\\ \gamma&\delta
\end{pmatrix}\begin{pmatrix}
1&h\\0&1
\end{pmatrix}$ and $U=\begin{pmatrix}
a&\widetilde{b}\\0&d
\end{pmatrix}$. This establishes the existence 
of $U$ and $S$ satisfying \refb{enon7}.

The analysis above has not yet
established that for given $r,s,t,u$, 
the choice of $S$ and $U$ is unique. To prove uniqueness,
suppose that $g=S'U'$ where
\begin{equation}
U'=\begin{pmatrix}
a'&b'\\0&d'
\end{pmatrix}~~~~\text{and}~~~~S'=\begin{pmatrix}
\alpha'&\beta'\\ \gamma'&\delta'
\end{pmatrix}\in SL(2,\ZZZ), \quad a',d'>0, \quad 0\le b'\le d'-1\, .
\end{equation}
Then $SU=S'U'$ tells us that $S^{-1} S'U'=U$. Using the form of $U$, $U'$, it
follows immediately that the lower left corner of $S^{-1}S'$ vanishes. Therefore,
$S^{-1}S'$ must have the form $\begin{pmatrix} 1 & k \cr 0 & 1\end{pmatrix}$ for
some $k\in \ZZZ$. This gives $a'=a$, $d'=d$, $b'=b-kd$. However since 
$0\le b,b'\le d-1$, $k$ must vanish and we have $b'=b$. This shows that $S'=S$
and $U'=U$.

We still need to check that the different choices of $a,b,d$ in \refb{enon7} always 
produce inequivalent $M(\AAA)$, i.e.\ there are no special matrices of the form $M_0$
for which 
\be \label{enon23}
U\, M_0 = S\,  U'  M_0, \quad U'\equiv
\begin{pmatrix} a' & b'\cr 0 & d'\end{pmatrix}\, ,
\ee
for some $SL(2,\ZZZ )$ matrix $S$. Since the quadratic forms associated
with the minor matrix are not identically zero, it follows from \refb{1} that at least
one matrix $\MM_{ij}$, obtained by keeping the 
$i$-th and the $j$-th
columns of $M_0$, must have non-vanishing determinant. Now,
it follows from \refb{enon23} that
\be\label{e318a}
U\, \MM_{ij} = S\, U'\, \MM_{ij}\, .
\ee
Since $\MM_{ij}$ has an inverse, we get
$U = S\, U'$.
However, 
due to the uniqueness of the decomposition given in
 \refb{enon7}, this is possible only for $U=U'$ and $S=I_2$ 
 as long as $a,b,d$ and $a',b',d'$ lie in the range
 given in \refb{enon7}.
 Therefore different choices of $(a,b,d)$  in \refb{enon7} 
produce inequivalent $M(\AAA)$.
 
There are some special quadratic forms for which we need to be
more careful. 
Let us suppose that the quadratic form \refb{ex1} and/or \refb{ex3} 
is such that it is
invariant under some $SL(2,\ZZZ)_S$ and/or $SL(2,\ZZZ)_U$ transformation\cite{buell}.
 For example the quadratic form $x^2+y^2$ is invariant under $\begin{pmatrix}
 0 & -1\cr 1 & 0\end{pmatrix}$. In that case two different charge vectors, related by
 this $SL(2,\ZZZ)_S$ and/or $SL(2,\ZZZ)_U$ transformation, will give same expressions
 for
 \refb{ex1} and \refb{ex3}. Therefore in order to determine if $U$ and $U'$ describe
 the same duality orbit, we could generalize \refb{enon23} by allowing this
$SL(2,\ZZZ)_S$ and/or $SL(2,\ZZZ)_U$   transformation to act on the right hand side.
 Now from the duality transformation laws \refb{e2}-\refb{e4}
 and the expression for the matrix $M(\AAA)$ given in \refb{enon2}, we see that an
 $SL(2,\ZZZ)_S\times SL(2,\ZZZ)_U$ transformation can be represented by a
 right multiplication of $M(\AAA)$ by some $4\times 4$ matrix $V$.\footnote{$V$ is actually
 an $SO(2,2;\ZZZ)$ matrix, but this will not be important for our analysis.}
 Therefore in order that $(a,b,d)$ and $(a',b',d')$ produce equivalent $M(\AAA)$,  
 it is enough to demand that
 \be \label{enon25}
U\, M_0 = S\,  U'  M_0 V \, ,
\ee
where $V$ represents some $SL(2,\ZZZ)_S\times SL(2,\ZZZ)_U$ transformation 
that leaves the quadratic forms \refb{ex1} and \refb{ex3} unchanged. Now since 
$M_0 V$ and $M_0$ generate the same quadratic forms, by Gauss's Lemma we have
\be \label{em0v}
M_0V=\wt S(M_0,V) M_0\, ,
\ee 
for some $SL(2,\ZZZ)$ matrix 
$\wt S(M_0,V)$ that depends on both $M_0$ and $V$. 
Substituting this into the right hand side of \refb{enon25}, we get
\be\label{enform}
U= S\, U' \, \wt S(M_0,V)\, ,
\ee
where we have used arguments similar to the one given below 
\refb{e318a}  to remove $M_0$ from both sides. 

As reviewed in Appendix \ref{smiddle}, the 
only special quadratic forms that are invariant under
some $SL(2,\ZZZ)$ transformation other than $\pm I_2$ 
are those proportional to $x^2+y^2$ or 
$x^2+xy+y^2$\cite{vol,buell}.\footnote{The $SL(2,\ZZZ)$ 
transformation $-I_2$ leaves
every quadratic form unchanged, but since the corresponding $\wt S(M_0,V)$ is
$-I_2$, it can be absorbed into the definition of $S$ in \refb{enform} and does not
affect $U'$. We need to be a bit more careful in the next section where $\Gamma_T$
represents a congruence subgroup of $SL(2,\ZZZ)$. If this subgroup does not include
$-I_2$, then we cannot absorb $-I_2$ into the definition of $S$ in \refb{enform}.
See footnote \ref{f17} for a discussion on this.}
It has been shown in Appendix \ref{sc} that in all cases, by suitably choosing
$M_0$, the matrix $\wt S(M_0,V)$ can be brought to one of the following three 
forms:
\be \label{elistmatrix}
\begin{pmatrix} 0 & -1 \cr 1 & 0\end{pmatrix}, \quad
\begin{pmatrix} 1 & -1 \cr 1 & 0\end{pmatrix}, \quad 
\begin{pmatrix} 0 & -1 \cr 1 & -1\end{pmatrix}\, .
\ee 
The first one is relevant when 
$q_1$ and/or $q_2$ is proportional to $(x^2+y^2)$, while the second and the
third ones are relevant when $q_1$ and/or $q_2$ is 
proportional to $(x^2+xy+y^2)$.

Let us now try to find solutions to \refb{enform}. For 
$\wt S(M_0,V)=\begin{pmatrix} 0 & -1 \cr 1 & 0\end{pmatrix}$, the equation takes
the form:
\be \label{efirstx}
\begin{pmatrix} a & b \cr 0 & d\end{pmatrix}=
\begin{pmatrix} \alpha & \beta \cr \gamma & \delta\end{pmatrix} 
\begin{pmatrix} a' & b' \cr 0 & d'\end{pmatrix} 
\begin{pmatrix} 0 & -1 \cr 1 & 0\end{pmatrix}\, , \quad 
\begin{pmatrix} \alpha & \beta \cr \gamma & \delta\end{pmatrix} \in SL(2,\ZZZ)\, .
\ee
This equation can be rewritten as
\be \label{erewrite}
\begin{pmatrix} a' & b' \cr 0 & d'\end{pmatrix} =
\begin{pmatrix} -\delta b + \beta d & \delta a \cr \gamma b -\alpha d & \
-\gamma a\end{pmatrix} \, .
\ee
Comparing the lower left corner, we get $\gamma b=\alpha d$. Since $\gamma$ and 
$\alpha$ are coprime, this can be solved by taking $\gamma=-d/\gcd(b,d)$,
$\alpha=-b/\gcd(b,d)$. 
We can now find $\delta$ and $\beta$ satisfying $\alpha\delta-\beta\gamma=1$,
but the result is determined only up to a shift symmetry: $\delta \to \delta + r\gamma$,
$\beta\to \beta+r\alpha$, $r\in\ZZZ$.
Comparing the lower right corners we now get $d'=
- \gamma a=ad / \gcd(b,d)$.  
Comparing the determinants on both sides we
get $a'=\gcd(b,d)$. Comparison of the upper left corners gives 
$-\delta b + \beta d =\gcd(b,d)$, but this is the same as the equation
$\alpha\delta-\beta\gamma=1$.
Comparison of the upper right corner gives  $b'=\delta a$. Due to the shift symmetry
mentioned above, $b'$ is determined up to an additive factor of $r\gamma a
= -r d'$, $r\in\ZZZ$. Using
this freedom, we can bring $b'$ in the range $0\le b'\le d'-1$. Therefore the final
results for $a', b', d'$ may be written as,
\be \label{efin11}
a'=\gcd(b,d), \quad d'=ad/\gcd(b,d), \quad b'\equiv \delta a \ \hbox{mod} \ d', \ 0\le b' < d'\, ,
\ee
where $(\delta,\beta)$ are any integer solutions to the equation
\be\label{efin12}
-\delta b + \beta d = \gcd(b,d)\, .
\ee
This tells us that when $q_1$ and/or $q_2$ is proportional to 
$x^2+y^2$, the orbits
labelled by $(a,b,d)$ and $(a',b',d')$ are the same if $(a',b',d')$ are related to
$(a,b,d)$ via \refb{efin11}, \refb{efin12}.

For 
$\wt S(M_0,V)=\begin{pmatrix} 1 & -1 \cr 1 & 0\end{pmatrix}$, the equation takes
the form:
\be \label{efirst1}
\begin{pmatrix} a & b \cr 0 & d\end{pmatrix}=
\begin{pmatrix} \alpha & \beta \cr \gamma & \delta\end{pmatrix} 
\begin{pmatrix} a' & b' \cr 0 & d'\end{pmatrix} 
\begin{pmatrix} 1 & -1 \cr 1 & 0\end{pmatrix}\, .
\ee
Equation \eqref{efirst1} can be written as:
\begin{equation}
\begin{pmatrix}
a'&b'\\0&d'
\end{pmatrix}=\begin{pmatrix}
-\delta b + \beta d&\delta(a+b)-\beta d\\\gamma b-\alpha d&-\gamma (a+b)+\alpha d
\end{pmatrix}\, .
\end{equation} 
This equation can be analyzed in the same way as \refb{erewrite} leading to the result:
\be\label{3aa}
a'=\gcd(b,d), \quad d'={ad\over \gcd(b,d)}, \quad b'=a'+\delta a, \quad
\beta d - \delta b =\gcd(b,d)\, .
\ee
The solution $(\beta,\delta)$ of the last equation has to be chosen so that $b'$ lies in
the range $(0,\ldots, d'-1)$.

For 
$\wt S(M_0,V)=\begin{pmatrix} 0 & -1 \cr 1 & -1\end{pmatrix}$, the equation takes
the form:
\be \label{esecond}
\begin{pmatrix} a & b \cr 0 & d\end{pmatrix}=
\begin{pmatrix} \alpha & \beta \cr \gamma & \delta\end{pmatrix} 
\begin{pmatrix} a' & b' \cr 0 & d'\end{pmatrix} 
\begin{pmatrix} 0 & -1 \cr 1 & -1\end{pmatrix}\, .
\ee
We can write this as
\be 
\begin{pmatrix} a' & b' \cr 0 & d'\end{pmatrix}= 
\begin{pmatrix} -\delta(a+b) +\beta d & \delta a \cr \gamma(a+b)-\alpha d & -\gamma a\end{pmatrix}\, .
\ee
This equation can also be analyzed in the same way as \refb{erewrite} leading to the result:
\be\label{e4de}
a'=\gcd(a+b,d), \quad d'={ad\over \gcd(a+b,d)}, \quad b'=\delta a, \quad
\beta d - \delta (a+b) =\gcd(a+b,d)\, .
\ee
Again, the solution $(\beta,\delta)$ of the last equation has to be chosen so 
that $b'$ lies in
the range $(0,\ldots, d'-1)$.

Therefore, when $q_1$ and/or $q_2$ is proportional to 
$x^2+xy+y^2$, the orbits
labelled by $(a,b,d)$ and $(a',b',d')$ are the same if $(a',b',d')$ are related to
$(a,b,d)$ via \refb{3aa} or \refb{e4de}.

\sectiono{Orbits of products of congruence subgroups of $SL(2,\ZZZ )$} \label{s3}

Let us now suppose that the actual duality group is $\Gamma_S\times\Gamma_T\times
\Gamma_U$ where $\Gamma_S$, $\Gamma_T$ and 
$\Gamma_U$ are congruence subgroups of $SL(2,\ZZZ )$. We shall allow them to be 
distinct
congruence subgroups. Our goal will be to classify the duality orbits of this theory.

The first task will be to classify the equivalence classes of binary
quadratic forms under the action
of a given congruence subgroup $\Gamma$
of $SL(2,\ZZZ )$. This problem was studied in \cite{1711.00230}.
Given a quadratic form $Ax^2+Bxy+Cy^2$, we consider the solution to the 
equation:\footnote{Note that due to the transformation law 
\refb{etrsxy} of $x,y$ under 
$\Gamma$, our variables $x$ and $y$ are related to those in \cite{1711.00230} by 
$x\leftrightarrow y$ exchange.}
\be
C\tau^2 + B\tau + A=0\, .
\ee
For $B^2<4AC$, this equation has a unique solution lying in the upper half plane:
\be\label{etau}
\tau=  \left\{-{B\over 2\, C} + {i\over 2\, |C|}\, \sqrt{4AC-B^2}\right\}\, .
\ee
We now choose a fundamental domain of $\Gamma$ in the upper half plane, and choose $A,B,C$ in the range so that $\tau$ lies inside the fundamental domain. 
For $\Gamma=SL(2,\ZZZ )$ this produces the range \refb{erange1}. 
For $\Gamma_0(p)$ with $p=2,3$, the restriction on $A,B,C$ 
may be chosen to be\cite{1711.00230}:
\be\label{erange2}
|B|\le |C|, \quad |B|\le p\, |A|, \quad AB>0 
\quad \hbox{for} \quad |B|=|C| \quad \hbox{or} 
\quad |B|=p\, |A|\, .
\ee
Similar results can be derived for other congruence subgroups of $SL(2,\ZZZ )$. 
It follows from the analysis of \cite{9807056,9807087,0401049} that $\tau$
given in \refb{etau} for the quadratic forms \refb{ex1}, \refb{ex3}, \refb{ex2} give the
attractor values of the $S$, $U$ and $T$ moduli respectively
of supersymmetric black holes of charge 
$(Q_1,\ldots, Q_4, P_1, \ldots, P_4)$, but this will not
play any role in our analysis.

Now suppose that we are given a charge vector and the associated Bhargava cube.
For this we can determine the $\Gamma_S$ and $\Gamma_U$ equivalence
classes of \refb{ex1} and \refb{ex3}, by making appropriate $\Gamma_S$ and 
$\Gamma_U$ transformations to bring the quadratic forms in the chosen range of 
values of $A,B,C$. 
This can be done irrespective of whether the quadratic forms are
primitive or not. The
question that we need to address now is: given the $\Gamma_S$ and 
$\Gamma_U$ equivalence
classes of \refb{ex1} and \refb{ex3}, what are the possible duality orbits?

Now, by following the analysis of \S\ref{s2} we can show that the
matrix $M(\AAA)$ constructed in \refb{enon2} can be expressed as 
in \refb{enon77}
for some reference matrix $M_0$ and an integer  
matrix $g$ of determinant $m$, with
$m$ defined as in \refb{enon88}. Since $\Gamma_T$ acts on $M(\AAA)$
by left multiplication, we see that different $g$, related by left multiplication by
elements of $\Gamma_T$, produce charge vectors in
the  same duality orbit. Therefore we can classify the independent duality orbits as
$\Gamma_T\backslash G$, where $G$ is the set of integer matrices of determinant
$m$. The solution to this problem may be found as follows. We have seen in 
\refb{enon7}
that any integer matrix of determinant $m$ can be expressed as the product of
an $SL(2,\ZZZ )$ matrix $S$ and a matrix $U$ of the form given in \refb{enon7}. On the
other hand, any $SL(2,\ZZZ )$ matrix can be expressed as an element of $\Gamma_T$ and
a representative element of the coset $\Gamma_T\backslash SL(2,\ZZZ )$. Therefore
the matrix $g$ can be expressed as
\be
S' \, \GG\, U, \quad S'\in\Gamma_T, \quad \GG\in \Gamma_T\backslash SL(2,\ZZZ ),
\quad U = \begin{pmatrix} a & b\cr 0 & d\end{pmatrix},
\ee
In that case the different possible choices of $\Gamma_T$ inequivalent 
$g$ appearing in the decomposition \refb{enon77} may be taken to be
\be\label{echoice}
\GG\, \begin{pmatrix} a & b\cr 0 & d\end{pmatrix}, \quad 
 \GG\in \Gamma_T\backslash SL(2,\ZZZ )\, .
 \ee
 
 In order to check that these always generate inequivalent duality orbits, we also 
need to check that we cannot have,
\be
\GG \, \begin{pmatrix} a & b\cr 0 & d\end{pmatrix} M_0= S' \,
\GG' \, \begin{pmatrix} a' & b'\cr 0 & d'\end{pmatrix} M_0\, , \quad
S'\in\Gamma_T, \quad \GG,\GG'\in \Gamma_T\backslash 
SL(2,\ZZZ ),
\ee
unless $S'$ is identify, $\GG'=\GG$ and 
$\begin{pmatrix} a & b\cr 0 & d\end{pmatrix} =
\begin{pmatrix} a' & b'\cr 0 & d'\end{pmatrix}$. To prove this we express this
equation as,
\be
\begin{pmatrix} a & b\cr 0 & d\end{pmatrix} M_0 =
\GG^{-1} S' \GG' \, \begin{pmatrix} a' & b'\cr 0 & d'\end{pmatrix} M_0\, .
\ee
Since $\GG^{-1} S'\GG\in SL(2,\ZZZ )$, it follows from the analysis below \refb{enon23}
that we must have $\begin{pmatrix} a & b\cr 0 & d\end{pmatrix} =
\begin{pmatrix} a' & b'\cr 0 & d'\end{pmatrix}$ and $\GG^{-1} S' \GG'$ must
be identity. Therefore $\GG=S' \GG'$. The uniqueness of the decomposition
of an $SL(2,\ZZZ )$ matrix into the product of a $\Gamma_T$ matrix and the 
representative of the coset $\Gamma_T\backslash SL(2,\ZZZ )$ then tells us that we
must have $\GG=\GG'$ and $S'=I_2$. This establishes that the orbits
associated with the different matrices of the form given in \refb{echoice} are distinct.
Therefore for a given choice of equivalences classes for the quadratic forms
\refb{ex1} and \refb{ex3}, represented by $q_1$ and $q_2$,
 the different duality orbits are labelled by minor matrices $M(\AAA)$ of the form:
 \be \label{efinlabalpre}
M(\AAA)=\GG\, \begin{pmatrix} a & b\cr 0 & d\end{pmatrix}\, M_0, \quad 
 \GG\in \Gamma_T\backslash SL(2,\ZZZ )\, .
 \ee
for some reference minor matrix $M_0$ associated with the Bhargava cube for which the
pair of quadratic forms are $q_1/m$ and $q_2/m$.

In special cases, when either $q_1$ is invariant under a subgroup of 
$\Gamma_S$ and/or $q_2$ is invariant under a subgroup of 
$\Gamma_U$, there may be further identification of the duality
orbits as in \S\ref{s2}. This means that $(\GG,a,b,d)$ and $(\GG',a',b',d')$ may
be equivalent for some special quadratic forms.
These special quadratic forms are not necessarily
those proportional to $(x^2+y^2)$ and $(x^2+xy+y^2)$, but are related to these
via $SL(2,\ZZZ)$ transformation.  For example $(x^2+y^2)$ is invariant under 
$\begin{pmatrix} 0 & -1\cr 1 & 0 \end{pmatrix}$, but this matrix may not be an element of
$\Gamma_S$ or $\Gamma_U$. However if we transform the quadratic form to
another by an $SL(2,\ZZZ)$ transformation matrix $K$ using \refb{etrsxy}, the
symmetry transformation that preserves the new quadratic form will be conjugated
by $K$ and may belong to $\Gamma_S$ or $\Gamma_U$ for suitable choice of $K$.
To see this let us represent $(x,y)$ as a row vector and let $q((x,y))$ be one
of the forms $x^2+y^2$ or $x^2+xy+y^2$ with symmetry generated by the $SL(2,\ZZZ)$
matrix $A$. Then we have $q((x,y)) = q((x,y)A)$. We now define $\tilde q((x,y))
= q((x,y)K)$ where $K\in SL(2,\ZZZ)$. Then we have,
\be
\tilde q((x,y)) = q((x,y)K) = q((x,y)KA) = \tilde q((x,y) KAK^{-1})\, .
\ee 
For generating inequivalent special quadratic forms we simply use for
$K$ all representative elements of the coset $\Gamma_S\backslash SL(2,\ZZZ)$
for $q_1$ and $\Gamma_U\backslash SL(2,\ZZZ)$ for $q_2$ and pick only those
candidates for which $KAK^{-1}$ belongs to $\Gamma_S$ and $\Gamma_U$
respectively.

Once we have identified the special $q_1$ and $q_2$ candidates, we can repeat
the analysis of \S\ref{s2} to find out which of the duality orbits, labelled by
$\GG,a,b,d$, are the same. We shall now show that the results of \S\ref{s2}
can be used to solve this problem. First consider the case where only one of $q_1$
or $q_2$ is special. Let us for definiteness suppose that $q_1$ is special. In
that case there are certain $\Gamma_S$ transformations which  leave $q_1$ invariant.
As in \S\ref{s2} we can represent this by some $4\times 4$ matrix $V$ 
multiplying $M_0$ from the
right and this, in turn, can be transformed into an $SL(2,\ZZZ)$ matrix $\wt S(M_0,V)$
multiplying  $M_0$ from the left. 
For
the
new Bhargava cube $\AAA'$, related to the original Bhargava cube $\AAA$ by a
duality transformation $V$ in $\Gamma_S$, \refb{efinlabalpre} is replaced by
\be\label{emodi}
M(\AAA')=\GG\, \begin{pmatrix} a & b\cr 0 & d\end{pmatrix}\, M_0V
=\GG\, \begin{pmatrix} a & b\cr 0 & d\end{pmatrix}\, \wt S(M_0,V)\, 
M_0, \quad 
 \GG\in \Gamma_T\backslash SL(2,\ZZZ )\, .
 \ee
Now, following the same arguments as in Appendix \ref{sc}, one can show that
$\wt S(M_0,V)$ either squares to $-I_2$ or cubes to $-I_2$.
and so, by appropriately redefining $M_0$, the matrix $\wt S(M_0,V)$
may be taken as one of the three forms given in \refb{elistmatrix}.\footnote{As in
\S\ref{s2}, we are not including in this list the matrix $-I_2$, or the matrices related to
\refb{elistmatrix} by a change in sign. However, if $-I_2$ is not an element of $\Gamma_T$,
then we have to include these matrices in our list, since we can no longer absorb $-I_2$
into $S$ in \refb{enform}. For example, if either $\Gamma_S$ or
$\Gamma_U$ includes $-I_2$ but $\Gamma_T$ does not include it, then, since every
quadratic form is invariant under $-I_2$, $\GG\in\Gamma_T\backslash
SL(2,\ZZZ)$ and  $-\GG\in\Gamma_T\backslash
SL(2,\ZZZ)$ will give the same duality orbit even though they represent different elements
of the coset.
\label{f17}}
 We can now follow the same procedure described in \refb{efirstx}-\refb{e4de} 
 to solve for
 $\alpha,\beta,\gamma,\delta,a',b',d'$ satisfying the equation:
 \be \label{e411}
 \begin{pmatrix} a & b\cr 0 & d\end{pmatrix}\, \wt S(M_0,V)=
 \begin{pmatrix} \alpha &\beta\cr \gamma & \delta \end{pmatrix} 
  \begin{pmatrix} a' & b'\cr 0 & d'\end{pmatrix}\, ,
  \ee
 and then represent $\GG \begin{pmatrix} \alpha &\beta\cr \gamma & \delta 
 \end{pmatrix}$ as a $\Gamma_T$ element multiplied by a possibly new element
 $\GG'$ of $\Gamma_T\backslash SL(2,\ZZZ)$. This will tell us that the orbits
 labelled by  $(\GG,a,b,d)$ and $(\GG',a',b',d')$ are the same.
 Note that even in the primitive case where $(a,b,d)=(1,0,1)$, this procedure may
 relate duality orbits labelled by different elements $\GG$ and $\GG'$ of
 $\Gamma_T\backslash SL(2,\ZZZ)$. We shall see an example of this below.
 
 Now consider the case where both $q_1$ and $q_2$ are special. 
 Then both $q_1$ and $q_2$ correspond to some $SL(2,\ZZZ)$ transformation
 of $m\, \bar q(x,y)$ where $m$ is an integer and
 $\bar q(x,y)$ is either $(x^2+y^2)$ or
 $(x^2+xy+y^2)$, but the $SL(2,\ZZZ)$ transformations may not be the same. In any case,
 regarding these $SL(2,\ZZZ)$ transformations as $SL(2,\ZZZ)_S$ and $SL(2,\ZZZ)_U$
 transformations, we can represent the combined transformation as 
 a multiplication of the minor matrix $M(\AAA)$ from the right by a $4\times 4$ 
  matrix $W\in SO(2,2;\ZZZ)$. 
 Therefore if $\bar M_0$ denotes a choice of the minor matrix for the
 pair of quadratic forms $(\bar q(x,y), \bar q(x,y))$,  
 then the minor matrix associated with the pair of quadratic forms
 $(q_1/m,q_2/m)$ can be taken to be
 \be
 M_0 =\bar M_0\, W\, .
 \ee
 Furthermore the $\Gamma_S\times \Gamma_U$ transformation under which
 $(q_1,q_2)$ is invariant will be given by $V=W^{-1}\bar VW$ where $\bar V$ 
is an $SL(2,\ZZZ)_S\times SL(2,\ZZZ)_U$ transformation that leaves
 invariant the pair of quadratic forms $(\bar q(x,y), \bar q(x,y))$.
 Therefore, we have
\be \label{econgr}
M_0 V = \bar M_0 \, W\, W^{-1} \bar V\, W = \wt S(\bar M_0, \bar V) \, \bar M_0 W
= \wt S(\bar M_0, \bar V) \, M_0\, ,
\ee
where in the third step we have used 
$\bar M_0\bar V = \wt S(\bar M_0, \bar V) \, \bar M_0$.
Therefore $\wt S(M_0,V) =\wt S(\bar M_0, \bar V)$. However we have seen
in Appendix \ref{sc} that
that $\wt S(\bar M_0, \bar V)$ is one of the three matrices appearing
in \refb{elistmatrix}. Therefore we can again use the procedure described in the 
previous paragraph to identify which sets of $(\GG,a,b,d)$ correspond to the
same duality orbit.

Since the special quadratic forms, for which we have
additional identification of orbits as described
above, are invariant under finite order elements of $\Gamma_S$ and/or $\Gamma_U$,
we shall recall a few results on finite order elements of congruence subgroups of
$SL(2,\ZZZ)$. It is known for example that $\Gamma(m)$ has no finite order elements
for $m\ge 2$, $\Gamma_1(m)$ has no finite order elements for $m\ge 4$ and 
$\Gamma_0(m)$ has no finite order elements if $-1$ and $-3$ are not squares modulo
$m$\cite{sebbar}. Therefore if $\Gamma_S$ and $\Gamma_U$ are picked from 
these congruence subgroups of $SL(2,\ZZZ)$, then we have no
special quadratic forms that are left invariant under any element of 
$\Gamma_S$ or $\Gamma_U$,
and we do not have any additional identification of duality orbits of the kind described
above. 

We shall now discuss some examples.
If $\Gamma_T=\Gamma_0(2)$, then the representative elements of the
coset $\Gamma_T\backslash SL(2,\ZZZ )$ may be taken as:
\be\label{eGamcoset}
\begin{pmatrix} 1 & 0\cr 0 & 1\end{pmatrix}, \quad 
\begin{pmatrix} 0 & 1\cr -1 & 0\end{pmatrix}, \quad 
\begin{pmatrix} 1 & 0\cr 1 & 1\end{pmatrix}\, .
\ee
In that case different possible choices of $g$ given in \refb{echoice} 
may be taken to be:
\be\label{elist}
\begin{pmatrix} a & b\cr 0 & d\end{pmatrix}, \quad
\begin{pmatrix} 0 & d\cr -a & -b\end{pmatrix}, \quad
\begin{pmatrix} a & b\cr a & b+d\end{pmatrix}\, , \quad ad=m, \quad 0\le b\le d-1\,.
\ee
We see that as long as $a\ne 0$, which in turn follows from $ad=m\ne 0$, the three
sets of matrices given in \refb{elist} are distinct. The minor matrix $M(\AAA)$
associated with different orbits take the form:
\be\label{efinlabel}
\begin{pmatrix} a & b\cr 0 & d\end{pmatrix} \, M_0, \quad
\begin{pmatrix} 0 & d\cr -a & -b\end{pmatrix}\, M_0, \quad
\begin{pmatrix} a & b\cr a & b+d\end{pmatrix}\, M_0\, , \quad ad=m, \quad 0\le b\le d-1\,.
\ee

If on the other hand, either $\Gamma_S$ or $\Gamma_U$ is $\Gamma_0(2)$,
then we have to classify the quadratic forms \refb{ex1} and/or \refb{ex3} by their
$\Gamma_0(2)$ equivalence classes, and also identify the special quadratic
forms for which there are additional identification of duality orbits.
The special quadratic forms for $\Gamma_0(2)$ can be found as follows. 
As discussed above,
they will be related to $x^2+y^2$ and $x^2+y^2+xy$ by appropriate elements of 
$SL(2,\ZZZ)$. To list the inequivalent ones, we can restrict our analysis to the
elements of the coset \refb{eGamcoset}. Consider first the quadratic form $x^2+y^2$
that is invariant under $\begin{pmatrix} 0 & -1\cr 1 & 0 \end{pmatrix}$. This matrix is
not in $\Gamma_0(2)$. It is easy to check that when we conjugate this by the second
element of \refb{eGamcoset}, the resulting matrix is still not in $\Gamma_0(2)$, but
when we conjugate this by the third
element of the coset \refb{eGamcoset} we get the matrix 
$\begin{pmatrix} 1 & -1\cr 2 & -1 \end{pmatrix}$ which is in $\Gamma_0(2)$.
The corresponding quadratic form, obtained by transforming $x^2+y^2$ by the third
element of \refb{eGamcoset}, is $x^2+2xy+2y^2$.
One can also check that the conjugation of 
$\begin{pmatrix} 1 & -1\cr 1 & 0 \end{pmatrix}$, which is the symmetry of 
$x^2+xy+y^2$, by any of the elements of the coset \refb{eGamcoset} does not
produce a
$\Gamma_0(2)$ matrix. Therefore we conclude that for $\Gamma_0(2)$, 
$x^2+2xy+2y^2$
is the only special quadratic form up to $\Gamma_0(2)$ transformations. 
It follows from our argument above that when
the quadratic forms \refb{ex1} and/or \refb{ex3} is proportional to $x^2+2xy+2y^2$,
then in the classification of the orbits by $(\GG,a,b,d)$ with 
$\GG\in \Gamma_T\backslash SL(2,\ZZZ)$, $a,b,d\in \ZZZ$, $0\le b\le d-1$, 
we need to identify $(\GG,a,b,d)$ and $(\GG',a',b',d')$, where the relation
between $a,b,d$ and $a',b',d'$ is given in \refb{efin11} and the relation 
between $\GG$
and $\GG'$ can be found by following the procedure described below
\refb{e411}.

We shall now give an example to illustrate how for the special quadratic forms,
the identification of the duality orbits labelled by $(\GG,a,b,d)$ and 
$(\GG',a',b',d')$ needs to be taken into account even for primitive quadratic forms
for which $(a,b,d)=(a',b',d')=(1,0,1)$. For this let us suppose that the duality group
is $SL(2,\ZZZ)_S\times \Gamma_0(2)_T\times SL(2,\ZZZ)_U$, and that the quadratic
form $q_1$ given in \refb{ex1}, transforming under $SL(2,\ZZZ)_S$, is 
proportional to $x^2+y^2$.
The choice of $q_2$ is arbitrary, but we take this to be coprime to $q_1$. 
Then we have
$a=1$, $b=0$, $d=1$ and for given $q_1$, $q_2$, the duality orbits are labelled by the
representative elements of the coset $\Gamma_T\backslash SL(2,\ZZZ)=
\Gamma_0(2)\backslash SL(2,\ZZZ)$. These have been listed in \refb{eGamcoset}.
We now apply \refb{e411} to determine which duality orbits should be identified. 
Since 
$\begin{pmatrix} a & b\cr 0 & d\end{pmatrix}$ is the identity matrix, we can get a
solution to \refb{e411} by setting
\be
\begin{pmatrix} \alpha & \beta\cr \gamma & \delta\end{pmatrix} = \wt S(M_0,V)
= \begin{pmatrix} 0 & -1\cr 1 & 0\end{pmatrix}, \quad
\begin{pmatrix} a' & b'\cr 0 & d'\end{pmatrix}= 
\begin{pmatrix} 1 & 0\cr 0 & 1\end{pmatrix}\, .
\ee
Therefore the new coset element $\GG'$ is given by
\be 
\GG' =\GG \, \begin{pmatrix} \alpha & \beta\cr \gamma & \delta\end{pmatrix} 
=\GG \begin{pmatrix} 0 & -1\cr 1 & 0\end{pmatrix}\, ,
\ee
up to multiplication by $\Gamma_0(2)$ matrices from the left. It is now easy to see
that this exchanges the first two coset representatives in 
\refb{eGamcoset}, leaving the third one unchanged. Therefore for 
a pair of coprime quadratic
forms, when one of the quadratic forms is proportional to $x^2+y^2$, we need to
identify the duality orbits labelled by first two elements of \refb{eGamcoset}.

The reader would have noticed that in our analysis we treat $\Gamma_S$ and 
$\Gamma_U$ differently from $\Gamma_T$. However since we are classifying the duality
orbits of $\Gamma_S\times \Gamma_T\times \Gamma_U$, the final result should be
symmetric in all three groups. We verify this in Appendix \ref{sd} by counting the number
of duality orbits in several examples.

So far we have assumed that the charge lattice is the lattice of integers. However in
specific string theory models, only a sublattice of the lattice of integers represents
allowed charges. In such cases many of the duality orbits described above will remain
empty.
For example for the STU model provided by example D of \cite{9508064},
the charge lattice
has the following 
form:
\be\label{erest1}
Q_1\in \ZZZ /2, \quad Q_2, Q_3, Q_4\in \ZZZ , \quad P_3\in 2\ZZZ , \quad P_1,P_2,P_4\in \ZZZ \, .
\ee
It can be checked that this lattice is invariant under 
$\Gamma_0(2)_S\times\Gamma_0(2)_T\times \Gamma_0(2)_U$ duality group 
with the
$\Gamma_0(2)$ actions as given in \refb{e2}-\refb{e4}, with $\alpha,\delta$ 
odd, $\gamma$ even and
$\beta$ arbitrary integer.  
However in this form the charge lattice is
not integral. 
We can rectify this by redefining the $Q_i$'s by mutiplying them
by a factor of 2. This does not
change the U and T-duality transformation rules given in \refb{e2} and \refb{e3}, but
the S-duality transformation given in \refb{e4} now becomes a $\Gamma^0(2)$ 
transformation. We can make this into a $\Gamma_0(2)$ transformation by 
relabelling $Q_i$'s as $P_i$'s and $P_i$'s as $Q_i$'s. Therefore the restriction on the
charges given in \refb{erest1} now translates to:
\be\label{erest2}
P_1\in \ZZZ , \quad P_2, P_3, P_4\in 2\ZZZ , \quad Q_3\in 2\ZZZ , \quad Q_1,Q_2,Q_4\in \ZZZ \, .
\ee
If we compute the quadratic
forms given in \refb{ex1}, \refb{ex3} and \refb{ex2} with these charges, then the 
coefficients of $y^2$ and $xy$ are always even. In our classification scheme,
this already eliminates all duality
orbits where one of these coefficients is odd, either for \refb{ex1} or for \refb{ex3}.
Furthermore, for a given choice of the reference matrix $M_0$, the choice of
$a,b,d$ in \refb{efinlabel} will be restricted so that $M(\AAA)$ given in \refb{efinlabel}
is compatible with the charge quantization condition after using \refb{enon2}.

We shall illustrate this with an example. It can be easily checked that given the
restriction \refb{erest2} on the charges, even and odd $P_1$ charges do not mix with
each other under $\Gamma_0(2)_S$, $\Gamma_0(2)_T$ and $\Gamma_0(2)_U$
transformations given in \refb{e2}-\refb{e4}. Let us consider the case where
$P_1\in 2\ZZZ +1$. In this case, it is easy to see that only
one of the three matrices in \refb{efinlabel} will satisfy the condition $P_1\in 2\ZZZ +1$,
$P_4\in 2\ZZZ $. For example let us suppose that the first matrix generates a charge
vector for which $P_1=2k+1$, $P_4=2\ell$ with $k,\ell\in \ZZZ $.
Then the second matrix in \refb{efinlabel}
will generate $P_1=2\ell$, $P_4=-2k-1$ and the third matrix in \refb{efinlabel}
will generate $P_1=(2k+1)$, $P_4=2k+2\ell+1$. Both of these violate the condition
$P_1\in 2\ZZZ +1$, $P_4\in 2\ZZZ $. Therefore for the 
subset of charges for which $P_1\in 2\ZZZ +1$,
the $\Gamma_0(2)$ equivalence classes of $q_1$, $q_2$ and the integers $a,b,d$
fix the duality orbit. Of course when one of the quadratic forms is special, 
we have further identification of duality orbits. For $P_1\in 2\ZZZ $
we do not have such simplifications and we need to include the orbits given by different
matrices in \refb{efinlabel}.

There are also other versions of the STU model with different charge 
lattice and the duality
group given by other congruence subgroups of $SL(2,\ZZZ )$.

In some cases, when two or all of the groups $\Gamma_S$, $\Gamma_T$ and
$\Gamma_U$ are identical congruence subgroups of $SL(2,\ZZZ)$,
the relevant duality group may also include some extra elements
that do not commute with 
$\Gamma_S$, $\Gamma_T$ and $\Gamma_U$, but permute them under
conjugation.
In such cases there will be additional identification among the duality orbits
classified above, that are
related by the action of these extra elements. We shall not discuss this,
but the details can be easily worked out. 
An example of this is provided in heterotic string theory compactified on $T^6$
if we consider an appropriately twisted index where the twist includes a change
of sign of four of the circles of $T^6$ and of the lattice associated with
the gauge group $E_8\times E_8$ or ${\rm Spin}(32)/\ZZZ_2$. In this case the only
charges that are invariant under this twist are the momentum and winding numbers
along the two remaining circles of $T^6$ and their magnetic counterparts.
The duality group that acts non-trivially on these charges is 
$SL(2,\ZZZ)_S\times O(2,2;\ZZZ)$ which can also be regarded as
$SL(2,\ZZZ)_S\times SL(2;\ZZZ)_T\times SL(2,\ZZZ)_U\rtimes \ZZZ_2$, where
the $\ZZZ_2$ exchanges $Q_1$ and $Q_3$. It can be seen from \refb{e2} and \refb{e3}
that this exchanges the $SL(2,\ZZZ)_U$ and $SL(2,\ZZZ)_T$ action on the charges.
Hence in this case, the duality orbit (for primitive quadratic forms) is specified by two equivalence classes, associated with the left-right and the front-back faces of the Bhargava cube shown in Fig.~\ref{figone}, with an additional equivalence relation that exchanges the two 
equivalence classes.

\bigskip

{\bf Acknowledgement:}
We wish to thank Shamit Kachru for comments on the manuscript and for
raising some interesting questions.
NB, SD and RKS would like to acknowledge the 
hospitality of HRI where part of the work was done. 
AS would like to  acknowledge the  hospitality of IISER Bhopal where this work began.
The work of NB is supported by SERB ECR grant 2019-2022. NB would like to 
thank Regular Associateship, the Abdus Salam ICTP, Trieste, Italy. 
The work of AB is supported by the grant no. MTR/2019/000582 from the SERB, Government of India.
The work of SD is supported by the grant no. EMR/2016/006294 and 
MTR/2019/000390 from the SERB, Government of India. SD also acknowledges 
the Simons Associateship of the Abdus Salam ICTP, Trieste, Italy.  
The work of AS was
supported in part by the J. C. Bose fellowship of 
the Department of Science and Technology, India and also by the Infosys 
Chair Professorship. Finally, we are grateful to people of India for their 
unconditional support towards researches in basic sciences.

\appendix

\sectiono{Proof of Gauss's Lemma} \label{sa}

In this appendix we shall describe the proof of Gauss's
Lemma following\cite{lemmermeyer}:
\begin{lemma}
Let 
\begin{equation}
M=\begin{pmatrix}
p_1&p_2&\dots &p_n\\
r_1&r_2 &\dots &r_n
\end{pmatrix}~~~
\text{and}~~~M'=\begin{pmatrix}
p_1'&p_2'&\dots &p_n'\\
r_1'&r_2' &\dots &r_n'
\end{pmatrix}
\end{equation}
be two $2 \times n$ integer matrices ($n \geq 3$) with the following properties:
\begin{enumerate}
\item the $2 \times 2-$minors of $M$ are coprime;
\item  there is an integer $m$ such that each minor of $M'$ is $m$ times the corresponding minor
of $M$. 
\end{enumerate}
Then there is a matrix $g =\begin{pmatrix}
a&b\\c&d
\end{pmatrix}$ with determinant $m$ and $a,b,c,d\in\ZZZ$ such that $M'= AM$.

\begin{proof}
Since the minors $M_{ik}=\begin{vmatrix}
p_i&p_k\\r_i&r_k
\end{vmatrix}=-M_{ki}$ are coprime, there exists $n(n-1)/2$ integers $x_{ik}$
for $i<k$ 
such that
\begin{equation}\label{exik}
\sum_{i,k=1\atop i<k}^{n} x_{ik}\begin{vmatrix}
p_i&p_k\\r_i&r_k
\end{vmatrix}=1\, .
\end{equation}
Now we have 
\begin{equation}\label{eaa3}
\begin{split}
p_k\begin{vmatrix}
p_i'&p_j'\\r_i&r_j
\end{vmatrix}+r_k\begin{vmatrix}
p_i&p_j\\p_i'&p_j'
\end{vmatrix}&=p_k p_i'r_j-p_kp_j'r_i+r_kp_ip_j'-r_kp_jp_i'\\&=p_i'(p_kr_j-r_kp_j)-p_j'(p_kr_i-p_ir_k)\\&={1\over m} \left\{p_i'(p_k'r_j'-r_k'p_j')-p_j'(p_k'r_i'-p_i'r_k')
\right\}
\\&=\frac{1}{m}p_k'\begin{vmatrix}
p_i'&p_j'\\r_i'&r_j'
\end{vmatrix}=p_k'\begin{vmatrix}
p_i&p_j\\r_i&r_j
\end{vmatrix}\, .
\end{split}
\end{equation}
Similarly, we have,
\be
p_k\begin{vmatrix}
r_i'&r_j'\\r_i&r_j
\end{vmatrix}+r_k\begin{vmatrix}
p_i&p_j\\r_i'&r_j'
\end{vmatrix}= r_k' \begin{vmatrix}
p_i&p_j\\r_i&r_j
\end{vmatrix}\, .
\ee
Let us define
\begin{equation}
\begin{split}
a=\sum_{i,j\atop i<j}x_{ij}\begin{vmatrix}
p_i'&p_j'\\r_i&r_j
\end{vmatrix},~~~~~b=\sum_{i,j\atop i<j}x_{ij}\begin{vmatrix}
p_i&p_j\\p_i'&p_j'
\end{vmatrix},\\
c=\sum_{i,j\atop i<j}x_{ij}\begin{vmatrix}
r_i'&r_j'\\r_i&r_j\end{vmatrix},
~~~~~d=\sum_{i,j\atop i<j}x_{ij}\begin{vmatrix}
p_i&p_j\\r_i'&r_j'
\end{vmatrix}\, .
\end{split}
\label{efirst}
\end{equation}
Then using \refb{exik}, \refb{eaa3}, \refb{efirst} we get,
\begin{equation}
ap_k+br_k=\sum_{i,j\atop i<j}x_{ij}p_k'\begin{vmatrix}
p_i&p_j\\r_i&r_j
\end{vmatrix}=p_k'\sum_{i,j\atop i<j}x_{ij}\begin{vmatrix}
p_i&p_j\\r_i&r_j
\end{vmatrix}=p_k'\, .
\end{equation}
Similarly we have $ cp_k + dr_k = r'_k $. Thus we have that 
\begin{equation}
\begin{pmatrix}
p_k'\\r_k'
\end{pmatrix}=g \begin{pmatrix}
p_k\\r_k
\end{pmatrix}, \quad g=\begin{pmatrix}
a&b\\c&d
\end{pmatrix}\, .
\end{equation} 
This gives $M'= gM$. Finally from the relation $M'_{ij}=det(g)M_{ij}$, obtained from the relation $M'= gM$, and hypothesis 2, we conclude that $det(g)=m$. 
\end{proof}
\end{lemma} 

\sectiono{Proof of existence of a Bhargava cube for a given pair of quadratic forms}
\label{sb}

In this appendix we shall prove, following \cite{lemmermeyer,trifkovic}, 
the existence of a Bhargava cube
for any given pair of quadratic forms of the same discriminant, even when they are
not primitive.

\theoremstyle{thm}
\begin{thm}
Let $q_i(x,y)=A_ix^2+B_ixy+C_iy^2,~~i=1,2$ be two quadratic forms of same discriminant $D$. Then there exists a Bhargava cube with $q_T=q_1$ and $q_L=q_2$, where $q_T$
is the quadratic form associated with the top-down faces of the cube and $q_L$ is the
quadratic form associated with the left-right faces of the cube.
\begin{proof}
\begin{figure}[H]
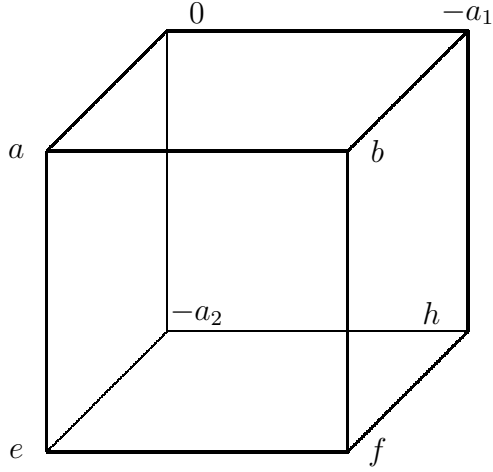

\centering
  \figonealt
  
\vskip .2in

  \caption{Bhargava cube for $q_1$ and $q_2$}
  \label{figonealt}
\end{figure}
We will construct a Bhargava cube of the form shown in Figure \ref{figonealt}.
The quadratic forms associated with the top-down and left-right faces of the cube
are given by: 
\begin{equation}\label{eb1}
\begin{split}
q_T(x,y) = (aa_1  )x^2 - (ah -a_1 e + a_2 b)xy - (eh + a_2 f)y^2, \\
q_L(x,y) = (aa_2  )x^2 - (ah + a_1 e -a_2  b)xy - (bh + a_1  f)y^2 .
\end{split}
\end{equation}
Equating the $x^2$ coefficients in $(q_1,q_2)$ with those in $(q_T,q_L)$, and
the $xy$ coefficient in $q_1+q_2$ with that in $q_T+q_L$,  we get,
\begin{equation}\label{eb11}
A_1=aa_1 ,~~~A_2=aa_2 ,~~~\frac{B_1+B_2}{2}=-ah. 
\end{equation}
Therefore we can solve \refb{eb11} by taking
\be \label{eb2}
a = \gcd(A_1 ,A_2 ,(B_1 + B_2 )/2), \quad a_1 =A_1/a, \quad 
a_2 =A_2/a, \quad h=-(B_1+B_2)/2a\, .
\ee  
This gives
\be \label{eb53}
\gcd(a_1,a_2,h)=1\, .
\ee
We note, however, that for $a>1$, other choices are also possible.
Equating the $y^2$ coefficients  in $(q_1,q_2)$ with those in $(q_T,q_L)$, we get 
\begin{equation}
he + a_2 f=-C_1,\qquad hb + a_1  f=-C_2.
\label{1'}
\end{equation} 
Finally, equating the coefficient of $xy$ in $q_1-q_2$ with that in $q_T-q_L$ we get,
\be \label{eb373}
a_1 e-a_2b ={B_1-B_2\over 2}\, .
\ee
Note that once \refb{eb373} is satisfied, only one of the two equations in 
\refb{1'} is independent. This is due to the fact that 
the since the discriminant of $q_1$
and $q_2$ are equal, once $A_1$, $A_2$, $B_1$ and $B_2$ have been fixed, $C_1$
is given in terms of $C_2$ and vice versa.

Equality of the discriminants of $q_1$ and $q_2$ give
\be\label{eb3}
B_1^2-4A_1C_1=B_2^2-4A_2C_2\implies {B_1-B_2\over 2} = {A_1C_1-A_2C_2
\over (B_1+B_2)/2} = - {(a_1C_1-a_2C_2)\over h}\, .
\ee
Using \refb{eb53} and \refb{eb3}, we get
\be\label{eb374}
\gcd(a_1,a_2)\bigg| {(B_1-B_2)\over 2}\, .
\ee
This in turn shows that \refb{eb373} admits integer solutions for $e$ and $b$. 
If $(e_0, b_0)$ denotes a particular pair of solutions, then the 
general solution takes the form:
\be\label{ebefin}
b = b_0+ r a_1 / \gcd(a_1,a_2), \quad e=e_0+ r a_2 / \gcd(a_1,a_2), \quad
r\in\ZZZ\, .
\ee
We now substitute this into \refb{1'} to find $f$. 
As discussed above, it is sufficient to
solve one of these equations, since the other will then follow by the equality
of the discriminants of $q_1$ and $q_2$. Therefore we can express the solution
as,
\be\label{eb375}
f = -{1\over a_2} \{C_1 + h e_0 + h a_2 r/ \gcd(a_1,a_2)\}, \quad \hbox{or}
\quad f = -{1\over a_1} \{C_2 + h b_0 + h a_1 r/ \gcd(a_1,a_2)\}\, .
\ee
The equality of the two solutions follows from the fact that $(e_0,b_0)$ solves
\refb{eb373}.  Let us define $q=\gcd(a_1,a_2)$.
Now for general choice of $r\in\ZZZ$, 
$f$ is not an integer. But from \refb{eb375} we can see that 
$a_2(f+hr/q)$ and $a_1(f+hr/q)$ 
are integers. Therefore, the denominator of $(f+hr/q)$ 
must divide $a_1$ and $a_2$ and hence must divide $q=\gcd(a_1,a_2)$. 
This allows us to write  $(f+hr/q)=p/q$ after appropriately multiplying the 
numerator and denominator of $f$ by an integer if necessary. This gives:
\be \label{effin}
f=  {p- h\, r\over \gcd(a_1,a_2)}, \quad p\in\ZZZ\, .
\ee
We now need to
find an $r$ that makes $f$ an integer. This is possible
if $\gcd(h, \gcd(a_1,a_2))=1$. The latter equation follows from \refb{eb53}.

This proves the existence of a Bhargava cube satisfying the constraints 
$q_T=q_1$ and $q_L=q_2$ for any pair of binary quadratic forms $q_1$ and
$q_2$.
\end{proof}
\end{thm}

Note that we 
nowhere assumed that the quadratic forms $q_1$ and/or $q_2$ are primitive. 
We have also not assumed that $q_1,q_2$ are coprime.
We shall end
this appendix by giving some examples of this construction for non-primitive $q_1,q_2$.
Instead of giving the Bhargava cube, we shall give the minor matrix associated with the
cube shown in Fig.~\ref{figonealt}:
\be \label{eminorb}
\begin{pmatrix} a & b & e & f\cr 0 & -a_1 & -a_2 & h \end{pmatrix}\, .
\ee
Consider the pair of quadratic forms:
\be
q_1=10(x^2+2xy+10y^2),\quad q_2=3x^2+300y^2\, .
\ee
Here neither $q_1$ nor $q_2$ is primitive, although $q_1$ and $q_2$ are
coprime. The corresponding minor matrix, constructed following the procedure
described in this appendix, is:
\be\label{emat3}
M(\mathcal{A})=\begin{pmatrix}
1&40&13&10\\0&-10&-3&-10
\end{pmatrix}\, .
\ee
Note that $M(\AAA)$ is not unique, but requires a choice of $r$ in \refb{ebefin},
\refb{effin},
which is determined up to shift by integer multiples of $\gcd(a_1,a_2)$.
If instead we consider the pair of quadratic forms:
\be
q_1=10m(x^2+2xy+10y^2),\quad q_2=m(3x^2+300y^2)\, ,
\ee
which are not even coprime,
we can still construct a
minor matrix by the procedure described 
in this appendix:
\be
\begin{pmatrix}
m&40m&13m&10m\\0&-10&-3&-10
\end{pmatrix}=
\begin{pmatrix}
m&0\cr 0 &1
\end{pmatrix} \begin{pmatrix}
1&40&13&10\\0&-10&-3&-10
\end{pmatrix}\, .
\ee
This is consistent with Gauss's Lemma.

As another example, consider the pair of quadratic forms:\be
q_1=4x^2+4xy+34y^2, \quad q_2=12x^2+12xy+14y^2\, .
\ee
Note that neither $q_1$ nor $q_2$ are primitive and they are also not coprime.
The corresponding minor matrix, constructed using the procedure described above, 
can be chosen to be:
\be \label{emat1}
M(\mathcal{A})=\begin{pmatrix}
4&7&17&0\\0&-1&-3&-2
\end{pmatrix} = \begin{pmatrix} 2 &1\cr 0 & 1\end{pmatrix}
\begin{pmatrix}
2&4&10&1\\0&-1&-3&-2
\end{pmatrix}\, .
\ee
The minor matrix $\begin{pmatrix}
2&4&10&1\\0&-1&-3&-2
\end{pmatrix}$ is the minor matrix associated with the quadratic forms 
$q_1/2$ and $q_2/2$. This is again consistent with
Gauss's Lemma.

Note that since the  minor matrix \refb{eminorb} 
always has a zero in its lower left corner, the matrix $g$ appearing in
\refb{enon99} will always have the form $\begin{pmatrix}
a &b\cr 0&d
\end{pmatrix}$ for some integers $a$, $b$ and $d$. Furthermore, since we have
chosen $a_1$, $a_2$ and $h$ to be coprime, $d$ must be 1. This of course is a 
particular choice. For a given pair of quadratic form, 
we can generate matrices with $d\ne 1$ by modifying our 
algorithm and allowing $a_1$, $a_2$ and $h$ to have a common factor.

\sectiono{Bhargava's results} \label{sbhargava}

In this appendix we shall briefly review the relevant parts of Bhargava's results that
we are using. Let $q_1$ and $q_2$ be a pair of primitive binary quadratic forms with the
same discriminant. Then it follows, as a special case of the results of appendix \ref{sb},
that there exists a Bhargava cube for which $q_1$ is the quadratic form associated
with the top-down faces of the cube, and $q_2$ is the quadratic form associated with the
left-right faces of the cube. Furthermore it follows from Gauss's Lemma with $m=1$ 
that if we 
consider two cubes with this property, then they are related by an 
$SL(2,\ZZZ)$ transformation acting on the entries connected by a front-back line. 
Therefore the quadratic form $q_3$ associated with the front-back face is determined 
uniquely up to an $SL(2,\ZZZ)$ transformation. This is the first part of the Bhargava's
theorem that we have used in our analysis in \S\ref{s1}. Of course, Bhargava's 
result says more -- that the product of the class group elements associated with the
quadratic forms $q_1$, $q_2$ and $q_3$ is the identity element of the class group, but
we have not used this result and will not discuss its proof.

The second part of Bhargava's results that we have used says
that two different charge vectors,
for which the corresponding Bhargava cubes have the same (primitive)
class group elements, can be
related to each other by a duality transformation. Let $\AAA$ and $\AAA'$ be the
Bhargava cubes associated with this pair of charge vectors, and let $(q_1,q_2,q_3)$
and $(q_1',q_2',q_3')$ be the corresponding quadratic forms. Since $q_1$ and
$q_1'$ describe the same class group elements, there is an $SL(2,\ZZZ)$ transformation
that relates $q_1$ to $q_1'$. Similar argument shows that there is an $SL(2,\ZZZ)$ 
transformation that relates $q_2$ to $q_2'$. We apply these transformations on the
entries of $\AAA$ connected by top-down and left-right lines respectively, producing
a new Bhargava cube $\AAA''$. The cubes $\AAA'$ 
and $\AAA''$ now have the same quadratic
forms associated with the top-down and left-right faces. Therefore, by 
Gauss's Lemma, there is an $SL(2,\ZZZ)$ transformation, acting on the entries
connected by front-back lines, that relate $\AAA''$ to $\AAA'$. Combining the two
duality transformations, one relating $\AAA$ to $\AAA''$ and the other 
relating $\AAA''$ to $\AAA'$, we get the 
duality transformation that relates $\AAA$ to $\AAA'$.

\newtheorem{Th}{Theorem}[section]

\sectiono{Automorphisms of binary quadratic forms} \label{smiddle}

In this appendix we shall review some basic facts about automorphisms of binary
quadratic forms (see \cite{vol,buell} for details).
We say that $K \in SL(2, \mathbb{Z})$ is a \emph{proper automorphism} of a binary quadratic form $f= Ax^2+Bxy+Cy^2$ if 
\be \label{emi1}
f(x, y)= f\left((x, y)K\right).
\ee
The group of proper automorphisms of $f$ is denoted by $\text{Aut}(f)$. 
We say that $f$ has a \emph{trivial} proper automorphism group if 
$\text{Aut}(f)= \{\pm I_2\}$, where $I_2$ is the $2\times 2$ identity matrix. 
Let $f=Ax^2+Bxy+Cy^2$ be a primitive binary quadratic form with discriminant $D$. Consider the associated Pell equation\cite{vol,buell}
\be\label{epell}
p^2- Dq^2= 4.
\ee
Given $(p, q) \in \mathbb{Z}^2$, we define
\be \label{emi2}
K(f, p, q)= \begin{pmatrix}
(p- q\, B)/2 & q\, A \\
-q\, C & (p+ q\, B)/2\end{pmatrix}\, .
\ee
It is straightforward to verify that $K$ given in \refb{emi2} satisfies \refb{emi1}.
We also see that for any $D$, $p=\pm 2$, $q=0$ is a solution to \refb{epell}, but
this generates the trivial automorphisms $\pm I_2$.
The following theorem gives a correspondence between solutions of Pell's equation and $\text{Aut}(f)$.
\begin{Th}
\label{mainthm}
The map $(p, q)\mapsto K(f, p, q)$ is a bijection between the set of integer solutions of $p^2- Dq^2= 4$ and $\mathrm{Aut}(f)$.
\end{Th}
The following results will also be useful.
\begin{Th}
\label{isom}
If $f$ and $g$ are primitive binary quadratic forms with the same discriminant, then $\text{Aut}(f)\cong \text{Aut}(g)$.
\end{Th}
\begin{Th}
If $f$ is a primitive positive definite quadratic form with discriminant $D \neq -3, -4$ then $\text{Aut}(f)$ is trivial. 
\end{Th}
This theorem tells us that the only cases we need to consider 
are $D= -3$ and $D= -4$. Let us introduce the matrices:
\be \label{edefW}
W_1=\begin{pmatrix}0 & -1\cr 1 & 0\end{pmatrix}, \qquad 
W_2=
\begin{pmatrix}1 & -1\cr 1 & 0\end{pmatrix}\, .
\ee
\begin{enumerate}
\item[(i)] If $D= -4$ then by Theorem \ref{isom}, it suffices to consider the
quadratic form $f_{-4}= x^2+y^2$.
Then we have $(p,q)=(\pm 2,0)$ or $(0,\pm 1)$, and,
\be\label{eCC4}
\text{Aut}(f_{-4})= \left\{ \pm I_2,\; \pm W_1\right\} = \left< W_1\right>\, ,
\ee
which is a cyclic group of order 4 generated by $W_1$. 
Note that the other inequivalent
quadratic form $-f_{-4}$ has the same automorphism group as $f_{-4}$.
\item[(ii)] Likewise, for $D= -3$ it suffices to consider the quadratic form
$f_{-3}= x^2+xy+y^2$. 
In this case we have $(p,q)=(\pm 2,0)$ or $(\pm 1, \pm 1)$, and,
\be\label{eCC6}
\text{Aut}(f_{-3})= \left< W_2\right>,
\ee
which is a cyclic group of order 6 generated by 
$W_2$.

\end{enumerate}

For non-primitive quadratic form $f$ with a common factor $r\in\ZZZ$, the
restriction on the automrphism groups can be found by considering the quadratic
form $f/r$ with discriminant $D/r^2$. The previous results tell us that Aut$(f)$ is
trivial unless $D=-3r^2$ or $-4r^2$, and is isomorphic to \refb{eCC4} for $D=-4r^2$
and \refb{eCC6} for $D=-3r^2$.

As described in \refb{etau},
there is also a correspondence between positive definite binary quadratic forms and points in the upper half plane. Given a quadratic form $f= Ax^2+Bxy+Cy^2$  
of discriminant $D$, the correspondence is
\be
f\mapsto -\frac{B}{2C} +  i\frac{\sqrt{|D|}}{2|C|}.
\ee
Using this, the form $x^2+y^2$ corresponds to the point $i$ and $x^2+xy+y^2$
corresponds to the point $\frac{-1+ i\sqrt{3}}{2}=e^{2\pi i/3}$.

One can use these results to restrict the possible automorphism groups of
quadratic forms inside congruence
subgroups. For example, for $D=-3$, we must have $B$ odd. Setting $B=2k+1$
with $k\in\ZZZ$, the equation $B^2-4AC=-3$ reduces to $AC=k(k+1)+1$. Therefore
$A$ and $C$ must be odd. Eq.\refb{emi2} now shows that since for non-trivial
automorphism $p=\pm1$ and $q=\pm 1$,
the off-diagonal elements of the matrix are both odd. This immediately rules out non-trivial 
Aut$(f)$ inside $\Gamma_0(N)$, $\Gamma^0(N)$, $\Gamma_1(N)$, $\Gamma^1(N)$
and $\Gamma(N)$ for even $N$ and $D=-3$.

\sectiono{Possible forms of $\wt S(M_0,V)$} \label{sc}

In this appendix we shall find the possible forms of $\wt S(M_0,V)$ introduced in
\refb{em0v}. Recall that $V$ represents an $SL(2,\ZZZ)_S\times SL(2,\ZZZ)_U$
transformation, represented by a $4\times 4$ matrix, that leaves some quadratic
form invariant. 
It follows from \refb{eCC4}, \refb{eCC6} that
in the range \refb{erange1}, the only examples of quadratic forms that are
invariant under some $SL(2,\ZZZ)$ transformation of the form given in 
\refb{etrsxy} are as follows\cite{buell}:\footnote{Via \refb{etau}, 
these are related to fixed points 
$i$ and
$e^{2\pi i/3}$ in the upper half plane under subgroups of $SL(2,\ZZZ)$.}
\ben\label{eform}
&& \hbox{$c(x^2+y^2)$  is invariant under $\mathrm{Aut}(f_{-4})$ generated by
$W_1$}, \nonumber \\
&& \hbox{$c(x^2+xy+y^2)$ is invariant under $\mathrm{Aut}(f_{-3})$ 
generated by $W_2$}\, ,
\een 
for any integer $c$. 
For a quadratic form proportional to $(x^2+y^2)$, the 
discriminant is $-4p^2$ for $p\in \ZZZ$ whereas for a quadratic form proportional
to $(x^2+xy+y^2)$ the discriminant is $-3q^2$ for $q\in \ZZZ$. Since we cannot have
$4p^2=3q^2$ for $p,q\in\ZZZ$, we see that we cannot have $q_1\propto (x^2+y^2)$
and $q_2\propto (x^2+xy+y^2)$ or vice versa. Therefore we have to consider the cases
where either only one of them is proportional to $(x^2+y^2)$ or $(x^2+xy+y^2)$ or
both of them are proportional to $(x^2+y^2)$ or both of them are proportional to
$(x^2+xy+y^2)$.

First consider the case where one of them is proportional to $(x^2+y^2)$. In this case,
$W_1$ given in \refb{edefW} (and its powers) is the symmetry of the quadratic form. 
This is represented
by a $4\times 4$ matrix $V$ multiplying $M_0$ from the right. The precise form
of $V$ depends on whether $W_1$ is in $SL(2,\ZZZ)_S$ or $SL(2,\ZZZ)_U$, i.e.\
whether $q_1$ or $q_2$ is proportional to $(x^2+y^2)$, but  since $W_1^2=-I_2$, we
must have $V^2=-I_4$ where $I_4$ is the $4\times 4$ identity 
matrix.\footnote{This follows from the fact that $-I_2$ in either $SL(2,\ZZZ)_S$,
$SL(2,\ZZZ)_T$ or $SL(2,\ZZZ)_U$ changes the signs of all the charges and
therefore changes the sign of the minor matrix. This is equivalent to multiplying the
minor matrix by $-I_4$ from the right.}
It then follows from \refb{em0v} that $\wt S(M_0,V)^2=-I_2$.
Therefore we must have (see {\it e.g.} page 55 of \cite{diamond})
\be
\wt S(M_0,V) = W^{-1} \, \begin{pmatrix} 0 & -1 \cr 1 & 0\end{pmatrix} W, 
\ee
for some $SL(2,\ZZZ)$ matrix $W$. In this case we can express \refb{em0v} as 
\be \label{em0v1}
W M_0 V = \begin{pmatrix} 0 & -1 \cr 1 & 0\end{pmatrix} \, W \, M_0\, .
\ee
If we now redefine $W M_0$ as the new $M_0$, then comparing \refb{em0v1} with
\refb{em0v} we get
\be\label{esim1}
\wt S(M_0, V) = \begin{pmatrix} 0 & -1 \cr 1 & 0\end{pmatrix}\, .
\ee

Next we consider the case where one of the quadratic forms is proportional to
$(x^2+xy+y^2)$. In this case,
$W_2$ given in \refb{edefW} (and its powers) is the symmetry of the quadratic form.
As in the previous case, let $V'$ be the $4\times 4$ matrix that represents 
this symmetry
by multiplying $M_0$ from the right.
Since $W_2^3=-I_2$, we must have $V^{\prime 3}=-I_4$. It follows from repeated application of
\refb{em0v} that $\wt S(M_0,V')^3 = -I_2$. Therefore either $\wt S(M_0,V')=-I_2$
or\cite{diamond}
\be\label{esim2}
\wt S(M_0,V') = \wt W^{-1} \, \begin{pmatrix} 1 & -1 \cr 1 & 0\end{pmatrix} \wt W ,
\ee
for some $SL(2,\ZZZ)$ matrix $\wt W$. We shall soon rule out the possibility that
$\wt S(M_0,V')=-I_2$. Therefore \refb{esim2} must hold. As before, we can redefine
$\wt W M_0$ as our new $M_0$, and get,
\be \label{esim3}
\wt S(M_0,V') = \begin{pmatrix} 1 & -1 \cr 1 & 0\end{pmatrix} \, .
\ee
It remains to show that we cannot have $S_0(M, V')=-I_2$. To prove this, assume
the contrary, so that \refb{em0v} gives
\be\label{esim4}
M_0 V'=- M_0 \, .
\ee 
Therefore the two rows of $M_0$ give two left eigenvectors of $V'$ of eigenvalue 
$-1$. However since $V'$ represents the action of $W_2$ on the charge vectors, this
will mean that the charge vectors have eigenvalue $-1$ under the action of $W_2$
-- as an element of $SL(2,Z)_S$ if $q_1\propto (x^2+xy+y^2)$ and as an
element of $SL(2,\ZZZ)_U$ for $q_2\propto (x^2+xy+y^2)$. However from the form
of $W_2$ given in \refb{edefW}, one can easily see that it does not have an eigenvector
with eigenvalue $-1$. Therefore we cannot have $\wt S(M_0,V')=-I_2$.

Next consider the case where both $q_1$ and $q_2$ are proportional to $(x^2+y^2)$.
Then they are invariant under $(W_1)_S$ and $(W_1)_U$ respectively. Let $V_1$ and
$V_2$ be the $4\times 4$ matrices, which, by multiplying $M_0$ from the right,
generate the actions of $(W_1)_S$ and $(W_1)_U$ respectively. Since 
$(W_1)_S$ and $(W_1)_U$ commute, it follows that $V_1$ and $V_2$ commute. Now,
following previous arguments, we can see that by appropriate choice of $M_0$, we can 
choose $\wt S(M_0,V_1)$ to be the matrix given in \refb{esim1}. 
Furthermore, since
$V_1$ and $V_2$ commute, it follows by repeated application of \refb{em0v} that
$\wt S(M_0,V_1)$ and $\wt S(M_0, V_2)$ also commute.
Also $\wt S(M_0,V_2)^2$ must be $-I_2$. It follows by explicit computation
that $\wt S(M_0,V_2)$ is $\pm \wt S(M_0,V_1)$. Therefore $\wt S(M_0,V)$, with 
$V$ given by arbitrary products of
powers of $V_1$ and $V_2$, will be given by the matrix
given in \refb{esim1} or its powers.

If both $q_1$ and $q_2$ are proportional to $(x^2+xy+y^2)$,
then they are invariant under $(W_2)_S$ and $(W_2)_U$ respectively. Let $V_3$
and $V_4$ be the $4\times 4$ matrices, which, by multiplying $M_0$ from the right,
generate the actions of  $(W_2)_S$ and
$(W_2)_U$ respectively. As before, we can choose $M_0$ appropriately to make
$\wt S(M_0,V_3)$ be given by the matrix appearing in \refb{esim3}. Since 
$(W_2)_S$ and $(W_2)_U$ commute, it follows that $V_3$ and $V_4$ commute and
also $\wt S(M_0,V_3)$ and $\wt S(M_0,V_4)$ commute. A short computation, using
the known form of $\wt S(M_0,V_3)$ given in \refb{esim3}, shows that $\wt S(M_0,V_4)$
must be given by some power of $\wt S(M_0,V_3)$. Therefore $\wt S(M_0,V)$, with
$V$ given by arbitrary products of powers of $V_3$ and $V_4$, will also be given by
powers of \refb{esim3}.

Since an overall minus sign in $V$ or $\wt S(M_0,V)$ can be absorbed into $S$
in \refb{enform}, by examining the matrices given in \refb{esim1}, \refb{esim3} 
and their powers, we see that 
we need to consider only three inequivalent choices of $\wt S(M_0,V)$:
\be \label{elistmatrixapp}
\begin{pmatrix} 0 & -1 \cr 1 & 0\end{pmatrix}, \quad
\begin{pmatrix} 1 & -1 \cr 1 & 0\end{pmatrix}, \quad 
\begin{pmatrix} 0 & -1 \cr 1 & -1\end{pmatrix}\, .
\ee 
The first one is relevant when 
either $q_1$ or $q_2$ is proportional to $(x^2+y^2)$, while the second and the
third ones are relevant when either $q_1$ or $q_2$ is proportional to $(x^2+xy+y^2)$.

\sectiono{Counting duality orbits in different ways} \label{sd}

In our analysis in the text, we have treated the duality group $\Gamma_T$ differently
from $\Gamma_S$ and $\Gamma_U$, in that we begin with a pair of 
quadratic forms representing equivalence 
classes of $\Gamma_S$ and $\Gamma_U$, and then find how many independent
Bhargava cubes exist that reproduce these quadratic forms and yet cannot be
related by $\Gamma_S\times \Gamma_T\times \Gamma_U$ transformations. 
However, we could exchange the roles of $\Gamma_S$ and $\Gamma_T$ or 
of $\Gamma_U$ and $\Gamma_T$ and carry out the counting. 
The
total number of duality orbits counted in these different ways must be the same. 
In this section we shall verify
this using several examples, We shall consider discriminants of the form $-4k^2$ and
$-3k^2$ with $k\in\ZZZ$, since for other discriminants there are no special quadratic
forms that remain invariant under an element of $SL(2,\ZZZ)$, and in such cases
the total number of duality orbits is given by the number of orbits of $SL(2,\ZZZ)^3$
times $n_Sn_Tn_U$ where $n_S$, $n_T$ and $n_U$ are the indices of $\Gamma_S$,
$\Gamma_T$ and $\Gamma_U$ in $SL(2,\ZZZ)$. This is manifestly symmetric under
permutations of $\Gamma_S$, $\Gamma_T$ and $\Gamma_U$. Also, to avoid 
repetition, we shall restrict the quadratic forms $q_1$ and $q_2$ to be positive definite.
If we allow $q_1$ and $q_2$ to be of either sign, the total number of orbits will be given
by four times what we shall count below, since the signs of $q_1$ and $q_2$ can be
changed independently.

\subsection{$D=-16$ and duality group $SL(2,\mathbb{Z})_S
\times\Gamma_0(2)_T\times SL(2,\mathbb{Z})_U$} \label{sd1}

\paragraph{Counting 1 :} We first carry out the analysis by letting $\Gamma_0(2)_T$
play the special role as in \S\ref{s3}. Therefore we begin by taking
pairs of quadratic forms representing 
equivalence classes of  $SL(2,\mathbb{Z})_S$ and $SL(2,\mathbb{Z})_U$. Possible inequivalent quadratic forms  for $D=-16$ are 
\ben\label{ed1}
f_1= x^2 + 4y^2 \quad \text{and} \quad f_2 = 2(x^2+y^2).
\een
Therefore, we have four possible pairs of $(q_1,q_2)$. 
We shall count the number of duality orbits for these four possible pairs.
Note that among the two quadratic forms in \refb{ed1}, 
$f_2$ is proportional to $f_{-4}$ and hence is invariant under $\mathrm{Aut}(f_{-4})$ 
which is generated by $W_1$.

\begin{itemize}
	\item
{\bf Pair 1, $q_1 = f_1$, $q_2 = f_1$ :} This pair is coprime, 
therefore the $U$ matrix introduced in \refb{enon7} is the identity matrix. 
The pair also does not contain any special form. Therefore number of duality orbits is
equal to number of representative elements of coset space $\Gamma_0(2)_T 
\backslash SL(2,\mathbb{Z})$ which is $3$.  Hence, for pair 1, we have 3 duality orbits.

\item 
{\bf Pair 2, $q_1 = f_1$, $q_2 = f_2$ :} This pair is also coprime, 
hence the $U$ matrix is again the identity matrix. 
However $q_2$ is a special form, and according to \refb{elistmatrix}, we have
 $\wt S(M_0, V) = \left( \begin{array}{cc} 0 & -1 \\ 1 & 0 \\ \end{array} \right)$. 
 Therefore coset elements $\GG$ and $\GG'$ 
 of $\Gamma_0(2)_T \backslash SL(2,\mathbb{Z})$, 
 related by $\mathcal{G}' = \mathcal{G} 
 \left( \begin{array}{cc} 0 & -1 \\ 1 & 0 \\ \end{array} \right)$, 
 describe the same duality orbit. 
 It is now easy to see that this exchanges the first two coset representatives 
 given in \refb{eGamcoset}, leaving the third one unchanged. Hence we have 2
 independent duality orbits for this pair.

\item
{\bf Pair 3, $q_1 = f_2$, $q_2 = f_1$:} The counting goes in the same way as for pair 2
and we get 2 independent duality orbits for this pair.

\item 
{\bf Pair 4, $q_1 = f_2$, $q_2 = f_2$ :} This pair is not coprime, since $m=\gcd(A_1,B_1,C_1,A_2,B_2,C_2) =2$. Hence the $U$ matrix defined in
\refb{enon7} can take three possible values: 
\be \label{eUlist}
\left( \begin{array}{cc} 1 & 0 \\ 0 & 2 \\ \end{array} \right), \quad 
\left( \begin{array}{cc} 1 & 1 \\ 0 & 2 \\ \end{array} \right), \quad
\left( \begin{array}{cc} 2 & 0 \\ 0 & 1 \\ \end{array} \right)\, . 
\ee
Each of these has three associated duality orbits from the three elements of the
coset $\Gamma_0(2)_T\backslash SL(2,\ZZZ)$. However there will be further
identification among these orbits 
since both the quadratic forms are special forms proportional to $f_{-4}$.
According to \refb{elistmatrix}, we have
 $\wt S(M_0, V) = \left( \begin{array}{cc} 0 & -1 \\ 1 & 0 \\ \end{array} \right)$. 
It is easy to see that the first matrix in \refb{eUlist}, multiplied by $\wt S(M_0,V)$ from
the right, gives the last matrix in \refb{eUlist}, multiplied by $-\wt S(M_0,V)$ from
the left. The effect of the latter is to further exchange the coset elements of 
$\Gamma_0(2)_T \backslash SL(2,\mathbb{Z})$ as in the case of pair 3. Therefore
the three duality orbits associated with the first matrix in \refb{eUlist} are
identified with the three duality orbits associated with the third matrix in \refb{eUlist},
and we get 3 independent orbits.
On the other hand the second element of \refb{eUlist}, multiplied by $\wt S(M_0,V)$ from
the right, gives us back the second element multiplied by $\begin{pmatrix} 1 & -1\cr 2 & -1
\end{pmatrix}$ from the left. Therefore we now need to check the effect of right
multiplication of the coset elements \refb{eGamcoset} by this matrix. It is easy to check
that this leaves invariant the first element of \refb{eGamcoset} and exchanges the 
second and the third elements. Therefore associated with the second element of
\refb{eUlist} there are 2 duality orbits. This gives a total of 5 duality orbits for this pair.

\end{itemize}

Adding up all the numbers, we get a total of 12 duality orbits.

\paragraph{Counting 2 :} Now we let $SL(2,\ZZZ)_U$ play the special role
that $\Gamma_0(2)_T$ played in the previous analysis.
For this we begin with the pairs of inequivalent quadratic forms of 
$SL(2,\mathbb{Z})_S$ and $\Gamma_0(2)_T$. 
The inequivalent quadratic forms of $SL(2,\ZZZ)$ are already listed in \refb{ed1}.
For $\Gamma_0(2)$ the inequivalent  quadratic forms are:
\ben \label{ed3}
	&& \tilde f_1 = x^2+4y^2, \qquad 
	\tilde f_2 = 2(x^2+y^2), \qquad
	\tilde f_3 = 4x^2+y^2, \nonumber \\ 
	&& \tilde f_4 = x^2+2xy+5y^2, \qquad
	\tilde f_5 = 2(x^2+2xy+2y^2) \, .
\een
Of these $\tilde f_5$ is a special form that is invariant under the order 4 element 
$\begin{pmatrix} 1 & -1\cr 2 & -1 \end{pmatrix}$
of $\Gamma_0(2)$.
Therefore we have total $10$ such pairs of $(q_1,q_2)$. We shall now count the number
of duality orbits associated with each such pair.
\begin{enumerate}
	\item 
The pairs $(f_1, \tilde f_1)$, $(f_1, \tilde f_2)$, $(f_1, \tilde f_3)$ and $(f_1, \tilde f_4)$ 
are coprime and do not contain any special form. Hence we have one duality orbit for each pair. This gives a total of $4$ duality orbits from these 4 pairs. 

\item 
$(f_1, \tilde f_5)$ pair is coprime and therefore $U$ is the identity matrix. 
Although it has one special form $\tilde f_5$, since the third duality group is $SL(2,\mathbb{Z})$, the special form plays no role and 
we have only $1$ duality orbit for this pair.  

\item 
The pairs $(f_2, \tilde f_1)$, $(f_2, \tilde f_3)$ and $(f_2, \tilde f_4)$ are also coprime. 
They contain one special form $f_2$, but since the third duality group 
is $SL(2,\mathbb{Z})$, there is only one orbit for each pair, Therefore we get 3 duality
orbits from these 3 pairs. 

\item 
Next consider the pair $(f_2, \tilde f_2)$.  
This pair is not coprime and has $m=2$ as a common factor. Therefore $U$ has one
of the three possible forms given in \refb{eUlist}. 
Furthermore $f_2$ is special, being invariant under an order four duality
transformation generated by $\begin{pmatrix} 0 & -1\cr 1 & 0\end{pmatrix}$.
It then follows from \refb{elistmatrix} that by 
choosing $M_0$ appropriately, the corresponding $\wt S(M_0,V)$ can be taken
to be $\begin{pmatrix} 0 & -1\cr 1 & 0\end{pmatrix}$. The analysis
described below \refb{eUlist} shows that the orbits associated with the first and the third
matrices in \refb{eUlist} must be identified. Therefore we have 2 duality orbits
associated with this pair.
\item 
Finally consider the pair $(f_2,\tilde f_5)$. This pair is not coprime and has $m=2$ 
as a common factor.  Furthermore both the forms are special forms,
being invariant under order four elements of the duality groups.
It follows from the arguments given below \refb{econgr} that by suitably choosing 
$M_0$, we can pick $\wt S(M_0,V)$ to be the matrix 
$\begin{pmatrix} 0 & -1\cr 1 & 0\end{pmatrix}$. The analysis now reduces to the previous
case, and gives a total of 2 duality orbits associated with this case.
\end{enumerate} 
Adding all the numbers, we get a total of 12 duality orbits. This agrees with the 
previous counting.

\subsection{$D=-16$ and duality group $\Gamma_0(2)_S \times SL(2,\mathbb{Z})_T
\times\Gamma_0(2)_U$ }

\paragraph{Counting 1 :}  Here we begin with a pair of quadratic forms 
$q_1$ and $q_2$ representing equivalence classes of 
$\Gamma_0(2)_S$ and $\Gamma_0(2)_U$, and classify the duality orbits for each such
pair. There are $5$ possible forms given in \refb{ed3}, 
hence $25$ possible pairs can be constructed.
\begin{enumerate}
		\item Out of these $25$ pairs $21$ pairs are coprime. 
		Since the third duality group is $SL(2,\mathbb{Z})$, these $21$ coprime pairs give $21$ orbits.
	\item There is one pair $(\tilde f_2, \tilde f_2)$ which is non-coprime 
	with $\gcd=2$ and none of the forms are special. 
	Therefore there are three $U$ matrices involved and hence this pair gives 
	$3$ orbits.
	\item There are two pairs $(\tilde f_2, \tilde f_5)$ and $(\tilde f_5, \tilde f_2)$ which 
	are non-coprime 
	with $\gcd=2$ and both the pairs contain one special form $\tilde f_5$ that is invariant
	under an order four element of $\Gamma_0(2)$. 
	Hence out of three $U$ matrices listed in \refb{eUlist},
	two will generate independent orbits. 
	Therefore, each pair gives $2$ orbits, and we get a total of $4$ orbits.
	\item Finally $(\tilde f_5,\tilde f_5)$ pair is non-coprime and both the forms are 
	special, being invariant under an order four subgroup of $\Gamma_0(2)$. 
	Hence out of three $U$ matrices two will be independent and hence this pair 
	will give $2$ orbits.
\end{enumerate}
Thus this way of counting gives a total of $30$ orbits.

\paragraph{Counting 2 :} We now exchange the roles of $SL(2,\ZZZ)_T$ and
$\Gamma_0(2)_U$. Therefore we begin with a pair of quadratic form 
($q_1,q_2$) representing 
equivalence classes of $SL(2,\mathbb{Z})_T\times\Gamma_0(2)_S$ and count the
number of independent duality orbits for each pair. Using \refb{ed1} and \refb{ed3}
we see that we have a total of 10 pairs of quadratic forms.

\begin{enumerate}
	\item Out of these ten pairs, four are coprime and contain no special 
	forms. Since the last duality group is $\Gamma_0(2)_U$ 
	these four pairs generate $4\times 3 =12$ orbits.
	\item $f_1, \tilde f_5$ are coprime but 
	$\tilde f_5$ is a special form, invariant under an order 
	four element of $\Gamma_0(2)$. 
	The corresponding $\wt S$ matrices will relate two of the three ${\cal G}$ 
	matrices listed in \refb{eGamcoset} to each other. 
	Hence this pair will generate $2$ orbits.
	\item $f_2$ pairing with $\tilde f_1, \ \tilde f_3$ and $\tilde f_4$ gives coprime 
	pairs but $f_2$ is a special form invariant under an order 4 element of 
	$SL(2,\ZZZ)$. Therefore each of these three pairs will generate $2$ orbits,
	giving a total of $6$ orbits.
	\item $(f_2,\tilde f_2)$ is non-coprime with $\gcd=2$ and has a special form. 
	This case is similar to that of {\bf Pair 4} in \S\ref{sd1}, and will generate $5$ orbits.
	\item $(f_2,\tilde f_5)$ is non-coprime with $\gcd=2$ and both the forms are 
	special forms. Again by arguments similar to 
	that of {\bf Pair 4} in \S\ref{sd1}, it will generate $5$ orbits. 
\end{enumerate}
Adding up all the numbers, we get a
total $30$ orbits. This agrees with the previous counting.

\subsection{$D=-36$ and duality group $SL(2,\mathbb{Z})_S\times SL(2,\mathbb{Z})_T\times\Gamma_0(2)_U$ }

In this case the inequivalent quadratic forms of $SL(2,\ZZZ)$ are 
\be \label{ed4}
f_1 = 9x^2+y^2, \qquad
f_2=2x^2+2xy+5y^2, \qquad
f_3=3(x^2+y^2)\, .
\ee
Of these $f_3$ is a special form, invariant under $W_1$.
On the other hand the inequivalent quadratic forms of  $\Gamma_0(2)$.
are:
\ben \label{ed5}
&& \hskip -.3in	\tilde f_1 = x^2+9y^2, \qquad
	\tilde f_2=x^2+2xy+10y^2, \qquad
	\tilde f_3=2x^2+2xy+5y^2, \qquad
	\tilde f_4= 2x^2-2xy+5y^2,\nonumber \\
&&  \hskip -.3in	\tilde f_5=3(x^2+y^2), \qquad
	\tilde f_6=3(x^2+2xy+2y^2), \qquad
	\tilde f_7 = 5x^2 +2xy +2y^2, \qquad
	\tilde f_8 =9x^2+y^2. \nonumber \\
\een
In this list, $\tilde f_6$ is a special form,  invariant under the order four element
$\begin{pmatrix} 1 & -1\cr 2 & -1 \end{pmatrix}$
of $\Gamma_0(2)$.

\paragraph{Counting 1 :} Here we proceed as usual by taking a pair of quadratic forms
$q_1$ and $q_2$, representing the equivalence classes of 
$SL(2,\mathbb{Z})_S$ and $\Gamma_0(2)_U$ respectively, and then find the number of
orbits for each pair. There are altogether 24 such pairs.

\begin{enumerate}
	\item There are 22 pairs of coprime forms. Each of these gives 1 orbit.
	Therefore we have 22 orbits from these 22 pairs. 
	\item  The pair $(f_3,\tilde f_5)$ is non-coprime  with $\gcd=3$ and 
	contains one special form $f_3$. There are four $U$ matrices, given by:
\be \label{eU3list}
\begin{pmatrix} 2 & 0\cr 0 & 1 \end{pmatrix}, \quad
\begin{pmatrix} 1 & 0\cr 0 & 2 \end{pmatrix}, \quad
\begin{pmatrix} 1 & 1\cr 0 & 2 \end{pmatrix}, \quad 
\begin{pmatrix} 1 & 2\cr 0 & 2 \end{pmatrix}\, .
\ee
It is easy to see that right multiplication of these matrices by
$\wt S=\begin{pmatrix} 0 & -1\cr 1 & 0 \end{pmatrix}$ 
relates the first two of these matrices to each other and the last two of these matrices
to each other, up to left multiplication by $SL(2,\ZZZ)$ matrices. Hence 
we have 2 independent $U$ matrices generating 2 orbits.
	\item There is one non-coprime pair $(f_3,\tilde f_6)$ with $\gcd=3$ and 
	both the forms are special. The same argument as above tells us that we have
	2 orbits for this pair.
\end{enumerate}
This gives a total of 26 orbits.

\paragraph{Counting 2 :} We now take a pair of quadratic forms $(q_1,q_2)$
representing equivalence classes of $SL(2,\ZZZ)_S$ and $SL(2,\ZZZ)_T$, and
count the duality orbits associated with each such pair. From \refb{ed4} we see that
we have  9 such pairs.

\begin{enumerate}
	\item We have 4 coprime pairs with no special forms. Each of these gives
	3 orbits associated with three elements of the coset $\Gamma_0(2)_U
	\backslash
	SL(2,Z)$ given in \refb{eGamcoset}. This gives  $12$ orbits.
	\item There are 4 coprime pairs with one special form. For each of these, the duality
	orbits generated by two of the three coset elements given in \refb{eGamcoset}
	get identified. Therefore we have $4\times 2=8$ orbits.
	\item We have one non-coprime pair $(f_3,f_3)$ with $\gcd=3$ and both 
forms special. There are four $U$ matrices listed in \refb{eU3list}, and we have
already seen that right multiplication by $\wt S$ exchanges the first two elements of
\refb{eU3list} with each other and the last two elements of \refb{eU3list} with each
other. Therefore, irrespective of how this acts on the elements of the coset
$\Gamma_0(2)_U\backslash SL(2,Z)$, the three orbits associated with the first
element of \refb{eU3list} gets identified with the three orbits associated with the second
element of \refb{eU3list}, and 
the three orbits associated with the third
element of \refb{eU3list} gets identified with the three orbits associated with the fourth
element of \refb{eU3list}. Therefore we have
6 orbits.
\end{enumerate}
Adding the numbers we get
26 orbits, in agreement with the previous counting.

\subsection{$D=-12$ and duality group $SL(2,\mathbb{Z})_S\times\Gamma_0(2)_T
\times SL(2,\mathbb{Z})_U$ }

In this case the  $SL(2,\mathbb{Z})$ inequivalent quadratic forms are 
\be\label{eq:form12SL}
f_1 = x^2+3 y^2, \qquad 
	f_2 = 2(x^2+xy +y^2).
\ee
$f_2$ is a special form that is invariant under $W_2$. On the
other hand, 
$\Gamma_0(2)$ inequivalent quadratic forms are:
\be\label{eq:form12Gamma}
	\tilde f_1 = x^2+3y^2, \qquad
	\tilde f_2=x^2+2xy+4y^2, \qquad
	\tilde f_3=2(x^2+xy+y^2), \qquad
	\tilde f_4= 3 x^2+y^2\, .
\ee
As discussed in \S\ref{s3}, there are no special forms among these.

\paragraph{Counting 1 :} We take a pair of quadratic forms $(q_1,q_2)$ representing
equivalence classes of $SL(2,\mathbb{Z})_S$ and $SL(2,\mathbb{Z})_U$ and
count the number of duality orbits for each such pair. There are four such pairs.

\begin{enumerate}
	\item $(f_1,f_1)$ : These are coprime and none of the forms are special.
Therefore the number of orbits is equal to number of elements of the coset
$\Gamma_0(2)_T\backslash SL(2,\mathbb{Z})$, which is 3.
	\item $(f_1,f_2)$ and $(f_2,f_1)$ : These are coprime and  one of the forms
	is special. Therefore we have two non-trivial $\wt S$ matrices given by the last
	two elements in \refb{elistmatrix}. They cyclically permute the three $\cal G$ 
	matrices listed in \refb{eGamcoset}. Thus each pair gives one orbit and
	we have 2 orbits in total.
	\item $(f_2,f_2)$ : These are non-coprime and have $\gcd =2$.
	Furthermore both of them are special forms. Therefore we have three non-trivial 
	$U$ matrices listed in \refb{eUlist}. It is straightforward to verify using \refb{3aa} and
	\refb{e4de} that the
	$\wt S$ matrices cyclically permute these three U matrices. Therefore, 
	we have a total of $3$ orbits coming from $3$ independent $\cal G$ matrices
	listed in \refb{eGamcoset}.
\end{enumerate}
This gives a total of  8 orbits.

\paragraph{Counting 2:} Here we pick $q_1,q_2$ to represent equivalence classes of
$SL(2,\mathbb{Z})_S$ and $\Gamma_0(2)_T$ respectively and count the number of
orbits associated with each pair.  We have 8 such pairs.
\begin{enumerate}
	\item There are four coprime pairs containing no special forms. These give 
	4 orbits.
	\item $f_2$ paired with $\tilde f_1,\ \tilde f_2$ and $\tilde f_4$ give coprime pairs
	but each pair contains one special form $f_2$. These give 3 orbits.
	\item $(f_2,\tilde f_3)$ is a non-coprime pair with $\gcd=2$ and one special 
	form $f_2$. The $\wt S$ matrices given by the last two elements in 
	\refb{elistmatrix} will cyclically permute the three $U$-matrices in \refb{eUlist}.
	Therefore we have  1 orbit.
\end{enumerate}
Adding up the numbers we get 8 orbits, in agreement with the previous result.

\subsection{$D=-12$ and duality group $SL(2,\mathbb{Z})_S\times\Gamma_0(2)_T
\times \Gamma_0(2)_U$ }

All the relevant quadratic 
forms are given in eqs.\refb{eq:form12SL}, \refb{eq:form12Gamma}. 

\paragraph{Counting 1 :} First we construct the pairs taking a
pair of quadratic forms representing 
equivalence classes $\Gamma_0(2)_U$ and $\Gamma_0(2)_T$. 
There are 16 such possible pairs with no special form.
\begin{enumerate}
	\item There are 15 coprime pairs and hence the $U$ matrix is identity. 
	Since the third duality group is $SL(2,\mathbb{Z})_S$, 
	these 15 coprime pairs give 15 orbits.
	\item There is only one non-coprime pair with $\gcd=2$. There are three possible $U$ matrices (\ref{eUlist}) associated with this pair. Since the pair does not have 
	any special form, the three $U$ matrices give independent orbits. 
	Hence this pair generates 3 orbits.
\end{enumerate}
Thus we get total 18 orbits.

\paragraph{Counting 2 :} We now take the first quadratic form of the pair from
equivalence 
classes of $SL(2,\mathbb{Z})_S$ and the second one from equivalence classes of
$\Gamma_0(2)_U$. Therefore there are 8 such pairs.

\begin{enumerate}
	\item There are 4 coprime pairs with no special forms. Since the third duality group is $\Gamma_0(2)_T$, there are three representative $\cal G$ matrices for each pair. Hence we get total 12 orbits.
	\item There are 3 coprime pairs with one special form. The special form generates two non-trivial $\wt S$ matrices, which cyclically permute the  
	three representative $\cal G$ matrices. Hence, each pair will generate 1 orbit and we have total 3 orbits from these three pairs.
	\item There is one non-coprime pair with $\gcd=2$ and containing one special form. Therefore, for this pair we have three possible $U$ matrices and two non-trivial $\wt S$ 
	matrices. These two $\wt S$ matrices will cyclically permute the 
	three $U$ matrices and hence this pair will generate 3 orbits due to three 
	different $\cal G$ matrices.
\end{enumerate}
Therefore we have total 18 orbits, in agreement with the previous counting.

\end{document}